\documentclass[runningheads,envcountsect,envcountsame]{llncs}
\usepackage[T1]{fontenc}
\catcode`\@=11
\spn@wtheorem{ldef}{Definition}{\bfseries}{\rmfamily}
\spn@wtheorem{lrem}{Remark}{\bfseries}{\rmfamily}
\catcode`@=12

\usepackage{reptheorem}
\usepackage{thmtools} 
\usepackage{thm-restate}

\usepackage{mathtools}
\usepackage{virginialake}
\vlnostructuressyntax
\vlnosmallleftlabels
\odframefalse
\odbackgroundtrue
\odbackgroundfirstfalse

\newcommand{\vlstr}[3]{\vltr{#1}{#2}{\vlhy{}}{#3}{\vlhy{}}}
\newcommand{\vlhtr}[2]{\vlstr{#1}{#2}{\vlhy{\hskip1.5em}}}

\newcommand{\proviso}[1]{\mbox{\small(#1)}}

\newcommand{\odf}[1]{\od{\odh{#1}}}

\newcommand{\cons}[1]{\{#1\}}
\newcommand{\conss}[1]{\left\{#1\right\}}
\newcommand{\conhole}{\cons\cdot}
\newcommand{\Chole}{C\conhole}
\newcommand{\Cphole}{C'\conhole}
\newcommand{\Ccon}[1]{C\cons{#1}}
\newcommand{\Cpcon}[1]{C'\cons{#1}}
\newcommand{\CCon}[1]{C\conss{#1}}

\newcommand{\Abot}[1][A]{#1^\bot}
\newcommand{\Cbhole}{C^\bot\conhole}
\newcommand{\Cbcon}[1]{C^\bot\cons{#1}}
\newcommand{\Cbpcon}[1]{C^{\prime\bot}\cons{#1}}

\newcommand{\prem}[1]{\mathsf{pr}(#1)}
\newcommand{\concl}[1]{\mathsf{cn}(#1)}

\newcommand{\rrule}{\rho}
\newcommand{\sysS}{\mathsf{X}}

\usepackage{newtxtext}       %
\usepackage[varvw]{newtxmath}       

\usepackage{hyperref}
\usepackage{cleveref}

\usepackage{comment}
\usepackage{graphicx}
\usepackage{color}

\usepackage{subfigure}

\newcommand{\defn}[1]{{\textit{\textbf{#1}}\/}}

\newcommand{\mqlap}[1]{\hbox to0pt{\hss$#1$\hss}}

\newcommand{\quiff}{\quad\mbox{iff}\quad}

\newcommand{\qualto}{\quad\leadsto\quad}

\newcommand{\atoms}{\mathcal{P}}
\newcommand{\atom}{p}
\newcommand{\atoma}{p}
\newcommand{\atomb}{q}
\newcommand{\agents}{\mathcal{A}}

\newcommand{\agenta}{a}
\newcommand{\agentb}{b}

\newcommand{\set}[1]{\{#1\}}

\newcommand{\tuple}[1]{\langle#1\rangle}

\newcommand{\aempty}{\varepsilon}
\newcommand{\powerset}[1]{\mathfrak{P}(#1)}
\newcommand{\erel}{\mathbin{\sim}}
\newcommand{\erelsym}{\mathord{\sim}}
\newcommand{\model}[1][]{\mathcal{M}_{#1}}

\newcommand{\validsym}{\mathord{\Vdash}}
\newcommand{\valid}[2][]{\validsym_{#1}\,{#2}}
\newcommand{\provesym}{\mathord{\vdash}}
\newcommand{\proves}[2][]{\provesym_{#1}\,{#2}}

\newcommand{\forcesym}{\mathord{\Vdash}}
\newcommand{\nforcesym}{\mathord{\not\Vdash}}
\newcommand{\forces}[3]{#1,#2\mathbin{\forcesym}#3}
\newcommand{\forcesM}[2]{\forces{\model}{#1}{#2}}
\newcommand{\forcesMw}[1]{\forcesM{w}{#1}}
\newcommand{\nforces}[3]{#1,#2\mathbin{\nforcesym}#3}
\newcommand{\nforcesM}[2]{\nforces{\model}{#1}{#2}}
\newcommand{\nforcesMw}[1]{\nforcesM{w}{#1}}

\newcommand{\intsym}{\tau}
\newcommand{\intof}[1]{#1^\intsym}
\newcommand{\iintof}[1]{(#1)^\intsym}

\newcommand{\OR}{\vee}
\newcommand{\AND}{\wedge}
\newcommand{\ANDOR}{\circledast}
\newcommand{\IMP}{\rightarrow}
\newcommand{\EQU}{\leftrightarrow}

\newcommand{\BOXDIA}{\bigcirc}
\newcommand{\DIA}{\Diamond}
\newcommand{\BOX}{\Box}

\newcommand{\BOXof}[1]{\BOX\left(#1\right)}

\newcommand{\PBOX}[1]{[#1]}
\newcommand{\PDIA}[1]{\langle#1\rangle}
\newcommand{\PBOXof}[2]{[#1]\left(#2\right)}

\newcommand{\cneg}[1]{\bar#1}
\newcommand{\cnegneg}[1]{\bar{\bar#1}}
\newcommand{\wcneg}[1]{\overline{#1}}

\newcommand{\ann}{\alpha}
\newcommand{\anna}{\alpha}
\newcommand{\annb}{\beta}
\newcommand{\acomp}{\cdot}

\newcommand{\deri}{\delta}

\newcommand{\derib}{\theta}
\newcommand{\deric}{\zeta}

\newcommand{\hderi}{\pi}

\newcommand{\hsq}[1][G]{\mathcal{#1}}
\newcommand{\hsqq}[1]{\mathcal{#1}}
\newcommand{\hsep}{\mathbin{\big\vert}}

\newcommand{\dsq}[1][\ann]{\mathbin{/\mkern-5mu/\mkern-3mu_{{#1}}}}
\newcommand{\dsqq}[1]{\mathbin{/\mkern-5mu/\mkern-3mu_{{#1}}}}

\newcommand{\mpr}{\mathrm{mp}}
\newcommand{\necr}{\mathrm{nec}}
\newcommand{\axr}{\mathrm{(Ax)}}
\newcommand{\topr}{\mathrm{(\top)}}
\newcommand{\orr}{\mathrm{(\OR)}}
\newcommand{\andr}{\mathrm{(\AND)}}
\newcommand{\cutr}{\mathrm{(cut)}}

\newcommand{\diatr}{\mathrm{(\DIA T)}}
\newcommand{\diafr}{\mathrm{(\DIA 5)}}
\newcommand{\diapr}{\mathrm{(\DIA P)}}
\newcommand{\boxr}{\mathrm{(\BOX)}}
\newcommand{\pboxzr}{\mathrm{(\PBOX{\cdot}_0)}}
\newcommand{\pboxr}{\mathrm{(\PBOX{\cdot}_1)}}
\newcommand{\pdiazr}{\mathrm{(\PDIA{\cdot}_0)}}
\newcommand{\pdiar}{\mathrm{(\PDIA{\cdot}_1)}}
\newcommand{\atr}{\mathrm{(Atom)}}
\newcommand{\newr}{\mathrm{(New)}}
\newcommand{\recallr}{\mathrm{(Recall)}}
\newcommand{\hpalr}{\mathrm{(Rule)}}

\newcommand{\Kax}{\mathsf{K}}
\newcommand{\Tax}{\mathsf{T}}
\newcommand{\Fax}{\mathsf{5}}
\newcommand{\Sfive}{\mathsf{S5}}
\newcommand{\Sfour}{\mathsf{S4}}
\newcommand{\PAL}{\mathsf{PAL}}
\newcommand{\DEL}{\mathsf{DEL}}
\newcommand{\EL}{\mathsf{EL}}
\newcommand{\IEL}{\mathsf{IEL}}

\newcommand{\HPAL}{\mathsf{HPAL}}
\newcommand{\OPAL}{\mathsf{OPAL}}

\newcommand{\IK}{\mathsf{IK}}
\newcommand{\IPAL}{\mathsf{IPAL}}

\newcommand{\MIPC}{\mathsf{MIPC}}

\renewcommand{\iD}{\mathsf{i}}
\renewcommand{\iU}{\mathsf{cut}}
\newcommand{\aiD}{\mathsf{ai}}
\newcommand{\kD}{\mathsf{k}}
\newcommand{\kpU}{\mathsf{k'}}
\newcommand{\fU}{\mathsf{5}}
\newcommand{\fD}{\mathsf{5'}}

\newcommand{\vD}{\mathsf{4'}}
\newcommand{\vU}{\mathsf{4}}
\newcommand{\tD}{\mathsf{t}}
\newcommand{\tU}{\mathsf{t'}}

\newcommand{\weakD}{\mathsf{w}}
\newcommand{\switchD}{\mathsf{s}}

\newcommand{\conD}{\mathsf{c}}
\newcommand{\necD}{\mathsf{e}}
\newcommand{\annD}{\mathsf{a}}

\newcommand{\aboxdot}{[\mkern-2mu\cdot\mkern-2mu]}
\newcommand{\diadot}{\langle\mkern-2mu\cdot\mkern-2mu\rangle}
\newcommand{\necPD}{\mathsf{e}_{\aboxdot}}
\newcommand{\kPD}{\mathsf{k}_{\aboxdot}}
\newcommand{\khPD}{\mathsf{k}^\ast_{\aboxdot}}
\newcommand{\kpPU}{\mathsf{k'_{\aboxdot}}}
\newcommand{\boxPD}{\mathsf{p_\BOX}}
\newcommand{\diaPD}{\mathsf{p_\DIA}}
\newcommand{\boxhPD}{\mathsf{p}^\ast_{\BOX}}
\newcommand{\diahPD}{\mathsf{p}^\ast_{\DIA}}

\newcommand{\funPD}{\mathsf{k_{\diadot}}}
\newcommand{\funhPD}{\mathsf{k^\ast_{\diadot}}}
\newcommand{\redD}{\mathsf{r}}
\newcommand{\negD}{\mathsf{n}}

\newcommand{\medD}{\mathsf{m}_\BOX}
\newcommand{\medvD}{\mathsf{m^4_\BOX}}
\newcommand{\medfD}{\mathsf{m^5_\BOX}}
\newcommand{\kfD}{\mathsf{5^\ast}}
\newcommand{\medPD}{\mathsf{m}_{\aboxdot}}
\newcommand{\medhPD}{\mathsf{m}^\ast_{\aboxdot}}
\newcommand{\kboxU}{\kpU}
\newcommand{\kpboxU}{\kpPU} 

\newcommand{\eboxD}{\mathsf{e_\BOX}}
\newcommand{\epboxD}{\mathsf{e_{\aboxdot}}}

\newcommand{\kPalD}{\mathsf{k[\cdot]}}

\newcommand{\weakr}{\mathrm{(w)}}
\newcommand{\extweakr}{\mathrm{(ew)}}
\newcommand{\liftr}{\mathrm{(lift)}}

\newcommand{\xMerge}{\oplus}
\newcommand{\ruler}{(\mathcal{R})}

\newcommand{\lift}{^{\uparrow C}}
\newcommand{\byIH}{\overset{\mathrm{(IH)}}{\rightsquigarrow} }

\newcommand{\action}{e} 
\newcommand{\actionb}{f} 
\newcommand{\actionM}{\mathcal{E}}
\newcommand{\Evt}{E}
\newcommand{\pre}{\mathrm{pre}}

\begin{document}
\title{Going Deep Inside Public Announcement Logic}
%
%
\author{Clara Lerouvillois\inst{1,2}\orcidID{0009-0008-8826-9050} \and
Lutz Stra{\ss}burger\inst{3,4}\orcidID{0000-0003-4661-6540}}
\authorrunning{C.~Lerouvillois and L.~Stra\ss burger}
%
\institute{IRIT, Université de Toulouse, France \and
IHPST, Université Paris 1 Panthéon Sorbonne, France
\and
LIX, Ecole Polytechnique, Institute Polytechnique de Paris, France
 \and
 Inria Saclay, France\\
 \url{https://www.lix.polytechnique.fr/Labo/Lutz.Strassburger/}
 }
\maketitle              
\begin{abstract}
    Dynamic Epistemic Logic (DEL) is a family of logics that extend Epistemic Logic by adding dynamic modalities to the language, which represent communication and yield updates of knowledge. Among its various systems, Public Announcement Logic (PAL) provides a simple and interesting framework to analyse how knowledge gets updated through public communication. While its semantics has been extensively studied, its proof-theory remains more fragmented. In particular, the structural analysis of the announcements and their composition is yet to be refined. Deep inference naturally appears as well suited proof formalism when it comes to decompose formulas from the inside. It also offers a formalism with less syntax, that dispenses with bureaucracy. This paper presents the first deep inference proof system for PAL, using the formalism of open deduction. We show soundness, completeness, and cut admissibility for the new system.
  
\end{abstract}
%
%


\section{Introduction}

Dynamic Epistemic Logic ($\DEL$) is a cohesive family of logical systems that represent \emph{knowledge} and its \emph{change} through \emph{communication}. $\DEL$ extends static Epistemic Logic ($\EL$), introduced by \cite{hintikka}, with additional \emph{dynamic} modalities used to represent communication events, or epistemic actions, such as public communication, semi-private communication to a group, private calls, etc. This means that, adding different such modalities yields different logical systems. Among them, the simplest and most studied is Public Announcement Logic ($\PAL$), introduced first in \cite{plaza1989} and then independently in \cite{gerbrandy1997reasoning}. $\PAL$ extends $\EL$ with a single dynamic modality $\PBOX \cdot $ that represents so-called \emph{public announcements}, where $\PBOX AB$ means that after any truthful announcement\footnote{Although announcements might be false in \cite{gerbrandy1997reasoning}, we follow the standard practice and consider only truthful announcements as in \cite{plaza1989}. Hence, an announcement is the public communication of a true formula to all agents.} of $A$, we have that $B$ holds.

Despite its apparent simplicity, $\PAL$ captures interesting properties of dynamics and knowledge, both at a conceptual and at a technical level, and has been used to solve famous problems like the Muddy Children puzzle \cite{DEL,ditmarsch_puzzles}.
Although $\PAL$ is as expressive as $\EL$, it is shown to be exponentially more succinct (see \cite{DEL}) while having the same computational complexity as $\EL$ (see \cite{lutz:PAL_complexity}), and therefore provides a more convenient and handy framework to study knowledge acquisition.
$\PAL$ has also been extended in various ways to capture several phenomena related to public communication in multi-agent systems, \emph{e.g.} considering arbitrary announcements \cite{balbiani2007:APAL,balbiani2008:APAL,baltag2023:APAL}, studying group announcements \cite{aagotnes2010group_announcement,wang2013:GAL,aagotnes2022:GAL} and group knowledge \cite{benthem2006logics_of_com_change,wang2011PAL+D} or investigating asynchronous announcements \cite{knight2019:AA,balbiani2022:AA,bal:dit:ler2025:AA}. Other AI-related applications of $\DEL$ are epistemic planning~\cite{bolander:andersen:11} and reasoning about knowledge in distributed systems~\cite{baltaga1998_PAL_and_DEL}.

The semantics of $\PAL$ has been extensively studied (see, \emph{e.g.}, \cite{plaza1989,baltaga1998_PAL_and_DEL,DEL}). \emph{Epistemic modalities} behave like $\Sfive$-modalities, \emph{i.e.}, each agent is associated with an equivalence relation, representing their uncertainty. Then, \emph{public announcements} act as \emph{model updaters} by restricting the set of possible worlds, eliminating all those which do not satisfy the formula being announced. When updating a model with announcement $A$, we therefore retain only the worlds which \emph{survive} the announcement. The epistemic model then progressively shrinks as successive announcements are made.

While the semantics is well-understood, the proof-theory of $\PAL$ is a bit more fragmented.
Its Hilbert-style axiomatisation (cf.~\cite{plaza1989,DEL,benthem2006logics_of_com_change})
extends that of multi-agent $\Sfive$ with reduction axioms, \textit{e.g.} $\PBOX Ap \EQU (A \IMP p)$, that allow to eliminate all dynamic modalities and translate any $\PAL$-formula into an equivalent static $\EL$-formula (that can be exponentially larger~\cite{DEL}).\footnote{Alternative axiomatisations have been proposed in \cite{wang2013axiomatizations}.} 
However, such an axiomatisation is of little use when it comes to finding proofs or studying their structural properties.
In the last decade, several analytic proof systems have been proposed for $\PAL$. Most of them either use inference rules that incorporate semantic reasoning on labels---this concerns the tableaux systems of~\cite{deBoerTableaux,balbiani2010tab} and the labelled sequent calculus of~\cite{nomura2015revising}---or use inference rules that simulate reduction axioms---as done by the labelled sequent calculus in~\cite{wu2023labelled} and the non-labelled sequent calculus in~\cite{liu2023}.
In order to obtain a label-free system that does not rely on reduction axioms, the work in~\cite{ler:pog:26} proposed \emph{dynamic hypersequents} that simulate the dynamics of announcements directly in the syntax.

In this paper, we propose an alternative route that is based on deep inference\cite{guglielmi2007,gug:str:01,brunnler:tiu:01}. In contrast to traditional proof formalisms that only allow \emph{shallow} inferences, deep inference allows applying inference rules at any depth inside a formula. It has been successfully employed for classical logic~\cite{brunnler:tiu:01,brunnler:06:herbrand,brunnler:06:locality,bru:gug:PC-DI}, intuitionistic logic~\cite{tiu:06:lpar,guenot:strassburger:14,gun:hei:par:LICS13}, various modal logics~\cite{hein:stewart:05,stouppa}, linear logic~\cite{str:02}, and many others, in particular, the logic $\BV$~\cite{guglielmi2007}, for which there cannot be a shallow inference system~\cite{tiu:SIS-II}.
Deep inference not only dispenses with heavy syntax and bureaucracy, providing more flexibility in the design of inference rules, it also gives alternative (and sometimes simpler) proofs of important theorems (\textit{e.g.} focusing~\cite{CGS:foccos} and interpolation~\cite{strassburger2026interpolation}).

In order to reduce the syntactic bureaucracy even more, we base our proof system $\OPAL$ on open deduction~\cite{GGP:open}. We then show soundness, completeness, and cut admissibility of this system via the hypersequent system of~\cite{ler:pog:26}, mentioned above. This makes the comparison between the two formalisms clear: hypersequent rules behave as huge inference steps that can be decomposed into smaller ones, represented by the deep inference rules.
To simplify the presentation, we confine ourselves here to the single-agent version of $\PAL$, as done in~\cite{ler:pog:26}.
Since $\PAL$ is conservative over $\Sfive$, our proof system is also a new deep inference proof system for $\Sfive$, which is surprisingly different from the one presented in~\cite{stouppa}.

The rest of the paper is organised as follows. \Cref{sec:PAL} introduces the syntax and semantics of $\PAL$, and \Cref{sec:HPAL} presents $\HPAL$, a one-sided version of the hypersequent calculus of \cite{ler:pog:26}. Then, in \Cref{sec:DI} we present our open deduction system $\OPAL$. Soundness and completeness are shown with respect to $\HPAL$ in \Cref{sec:soundness} and \Cref{sec:completeness}, respectively. Finally, in \Cref{sec:cut} we show that the cut is admissible in $\OPAL$, by relying on the corresponding result for $\HPAL$.


\section{Syntax and Semantics of $\PAL$}
\label{sec:PAL}

We give a brief introduction to the syntax, semantics and Hilbert-style axiomatisation of the single-agent fragment of Public Announcement Logic ($\PAL$). This is a mere adaptation of the standard presentation as is can be found, for example, in~\cite{DEL,ler:pog:26}.

The \defn{formulas} 
are generated from a countable set $\atoms=\set{\atoma,\atomb,\ldots}$ of \defn{atoms} via the following grammar:
\begin{equation}
  \label{eq:formula}
  A,B\coloneqq \top\mid\bot\mid \atom\mid\cneg\atom\mid
  A\AND B\mid A\OR B\mid\BOX A\mid\DIA A\mid\PBOX AB\mid\PDIA AB
\end{equation}
where $\top$ and $\bot$ are \defn{units}, and $\BOX$ and $\DIA$ are called \defn{epistemic modalities}, and $\PBOX\cdot$ and $\PDIA\cdot$ are called \defn{announcement modalities}.
We consider formulas only in \emph{negation normal form} to simplify the presentation and to ease the readability of the proof systems in this paper. Then,
\defn{negation} is defined for all formulas via De~Morgan dualities:
\begin{equation}
  \label{eq:deMorgan}
  \cnegneg \atom\;=\;\atom
  \qquad
  \begin{array}{r@{\;=\;}l@{\qquad}r@{\;=\;}l@{\qquad}r@{\;=\;}l@{\qquad}r@{\;=\;}l@{\qquad}}
    \cneg\top&\bot&
    \wcneg{A\AND B}&
    \cneg A\OR\cneg B&
    \wcneg{\BOX A}&\DIA \cneg A&
    \wcneg{\PBOX AB}&\PDIA A\cneg B
    \\[.5ex]
    \cneg\bot&\top&
    \wcneg{A\OR B}&
    \cneg A\AND\cneg B&
    \wcneg{\DIA A}&\BOX \cneg A&
    \wcneg{\PDIA AB}&\PBOX A\cneg B
  \end{array}
\end{equation}
It follows that for all formulas $\cnegneg A=A$, and we define \defn{implication} as $A\IMP B=\cneg A\OR B$ and \defn{equivalence} as $A\EQU B=(A\IMP B)\AND(B\IMP A)$.

The \defn{axioms} of $\PAL$ are the axioms of classical propositional logic, together with the following axioms (where $p$ can be any atom, negated atom, or unit):
\begin{equation}
  \label{eq:axioms}
  \begin{array}{l}
    \BOX(A\IMP B)\IMP(\BOX A\IMP \BOX B)\\
    \BOX A\IMP A\\
    \DIA A\IMP\BOX\DIA A
  \end{array}
  \hskip4em
  \begin{array}{l}
    \PBOX A\atom\EQU (A\IMP \atom)\\
    \PBOX AB\EQU (A\IMP \PDIA AB)\\
    \PBOX A(B\AND C)\EQU(\PBOX AB\AND \PBOX AC)\\
    \PBOX A\BOX B\EQU (A\IMP \BOX\PBOX AB)\\
    \PBOX A\PBOX B C\EQU\PBOX{A\AND \PBOX AB}C
  \end{array}
\end{equation}
where the axioms on the left are the called $\Kax$, $\Tax$, and $\Fax$, respectively, and they define the classical modal logic~$\Sfive$. The axioms on the right in~\eqref{eq:axioms} are called \defn{reduction axioms}.\footnote{Our axiom $\PBOX AB\EQU (A\IMP \PDIA AB)$ is equivalent to  $\PBOX A\cneg B\EQU (A\IMP \wcneg{\PBOX AB})$ given in~\cite{ler:pog:26}.}
The \defn{theorems} of the logic $\PAL$ are the formulas that can be derived from the axioms in~\eqref{eq:axioms} using the rules of \defn{modus pones} and \defn{necessitation}, shown below:
\begin{equation}
  \label{eq:mp}
  \vliinf{\mpr}{}{B}{A}{A\IMP B}
  \hskip6em
  \vlinf{\necr}{}{\BOX A}{A}
\end{equation}
We write $\proves[\PAL]A$ iff $A$ is a theorem of $\PAL$.

\begin{lemma}\label{lem:reduction}
  The following formulas are theorems of\/ $\PAL:$
  \begin{equation}
    \label{eq:reduction2}
    \begin{array}{l}
      \PDIA A\atom\EQU (A\AND\atom)\\
      \PDIA AB\EQU (A\AND \PBOX AB)\\
      \PDIA A(B\AND C)\EQU(\PDIA AB\AND \PDIA AC)\\
      \PDIA A\BOX B\EQU (A\AND \BOX\PBOX AB)\\
      \PDIA A \PDIA B C \EQU \PDIA {\PDIA AB}C
    \end{array}
    \hskip4em
    \begin{array}{l}
      \PDIA A(B\OR C)\EQU(\PDIA AB\OR \PDIA AC)\\
      \PDIA A \DIA B \EQU A \AND \DIA \PDIA AB\\
      \PDIA A \PBOX BC \EQU (A \AND \PBOX {\PDIA AB} C)\\[.5ex]
      \PBOX A(B\OR C)\EQU(\PBOX AB\OR \PBOX AC)\\
      \PBOX A \DIA B \EQU( A \IMP \DIA \PDIA AB)\\
      \PBOX A \PDIA BC \EQU (A \IMP \PDIA{\PDIA AB}C)
    \end{array}
  \end{equation}
\end{lemma}
\begin{proof}
  They all follow by straightforward derivations.
  \qed
\end{proof}

\begin{theorem}
  For every formula $A$ there is a formula $A'$ with $\proves[\PAL]{A\EQU A'}$ such that $A'$ does not contain any announcement modalities.
\end{theorem}

\begin{proof}
  This can be shown by induction on $A$, by repeatedly applying \Cref{lem:reduction}.
  \qed
\end{proof}

Let us now turn to the semantics of $\PAL$, which is, as expected, based on the standard Kripke semantics for modal logics~\cite{kripke1959}.

\begin{definition}
  An \defn{(epistemic) model} is a tuple $\model=\tuple{W,\erel,V}$ where $W$ is a set of possible \defn{worlds}, and $\erelsym\subseteq W\times W$ is an equivalence relation (\emph{i.e.}, transitive, symmetric, and reflexive), called the \defn{accessibility relation}, and $V\colon W\to\powerset{\atoms}$ is the \defn{valuation function} mapping each world $w\in W$ to the set of atoms that are true at $w$.
\end{definition}

\begin{definition}
  Let $\model=\tuple{W,\erel,V}$ be an epistemic model. We now define the \defn{forcing relation} $\forcesMw A$ and the \defn{updated model} $\model[A]=\tuple{W_A,\erel_A,V_A}$ simultaneously by induction on $A$, as shown below:
  \begin{equation}
    \label{eq:forcing}
    \begin{array}{l@{\quiff}l}
      \forcesMw \top & \mbox{always}\\
      \forcesMw \atom & \atom\in V(w)\\
      \forcesMw \cneg\atom & \atom\notin V(w)\\
      \forcesMw A\AND B & \forcesMw A \mbox{~and~} \forcesMw B\\
      \forcesMw A\OR B & \forcesMw A \mbox{~or~} \forcesMw B\\
      \forcesMw \BOX A & \mbox{for all $v$ with $w\erel v$ we have~}\forcesM v A\\
      \forcesMw \DIA A & \mbox{there is a $v$ with $w\erel v$ such that~}\forcesM v A\\
      \forcesMw \PBOX AB & \mbox{if~}\forcesMw A \mbox{~then~}\forces{\model[A]}wB\\
      \forcesMw \PDIA AB & \forcesMw A \mbox{~and~} \forces{\model[A]}wB
    \end{array}
  \end{equation}
  and
  \begin{equation}
    \label{eq:update}
    \begin{array}{r@{\;\;=\;\;}l}
      W_A& \set{w\in W\mid \forcesMw A}\\
      \erel_A&\erelsym\cap(W_A\times W_A)\\
      V_A(w)&V(w)\qquad\mbox{(for all $w\in W_A$)}
    \end{array}
    \hskip4em
  \end{equation}
  We write $\nforcesMw A$ iff $\forcesMw A$ does not hold. In particular, we always have  $\nforcesMw \bot$ and it follows that $\forcesMw \cneg A$ iff $\nforcesMw A$, for all formulas $A$. We say that $A$ is \defn{valid}, denoted as $\valid[\PAL]A$, iff we have $\forcesMw A$ for every model $\model$ and every world $w$ in that model.
\end{definition}

\begin{remark}\label{rem:nec-pbox}
  Observerve that from this definition it follows that for all formulas $A$ and~$B$, we have that $\valid[\PAL] B$ implies $\valid[\PAL] {\PBOX AB}$.
\end{remark}
The following has been shown in \emph{e.g.}~\cite{DEL}.
\begin{theorem}\label{thm:PAL}
  For all formulas $A$, we have that\/ $\proves[\PAL] A$ iff\/ $\valid[\PAL]A$.
\end{theorem}


\section{Hypersequents for $\PAL$}\label{sec:HPAL}

We now present a one-sided version of the hypersequent calculus from \cite{ler:pog:26}.
Here, a \defn{(one-sided) sequent} $\Gamma$ is a finite, possibly empty, multiset of formulas, separated by comma, \emph{i.e.}, a sequent is written as $A_1,A_2,\ldots,A_k$ for some $k\ge0$. An \defn{announcement} $\ann$ is a finite, possibly empty list of formulas, separated by a dot, \emph{i.e.},  an announcement is written as $A_1 \cdot A_2\cdot\ldots\cdot A_h$ for some $h\ge0$. For an announcement $\ann$ and a formula $B$, we define the formulas $\PBOX \ann B$ and $\PDIA \ann B$ inductively as follows:
\begin{equation}
  \label{eq:ann}
  \begin{array}{r@{\;\;:=\;\;}l@{\qquad}r@{\;\;:=\;\;}l}
    \PBOX\aempty B & B & \PDIA\aempty B & B \\
    \PBOX{\ann \cdot A} B & \PBOX{\ann}{\PBOX A B}  &
    \PDIA{\ann \cdot A} B & \PDIA{\ann}{\PDIA A B}    
  \end{array}
\end{equation}
  where $\aempty$ is the empty list of formulas.

\begin{definition}
  A \defn{dynamic sequent} $X$ is an expression $\dsq[\ann_1]\Gamma_1\dsq[\ann_2]\Gamma_2\cdots\dsq[\ann_n]\Gamma_n$ for some $n\ge1$ where $\ann_1,\ldots,\ann_n$ are announcements and $\Gamma_1,\ldots,\Gamma_n$ are sequents, and where $\ann_i\neq\ann_j$ for all $1\le i<j\le n$. For $\ann=\aempty$ we write $\Gamma$ for $\dsq[\aempty]\Gamma$. In particular, then every sequent $\Gamma$ is a dynamic sequent.
  If $X_1,\ldots,X_m$ for some $m\ge1$ are dynamic sequents, then $\hsq=X_1\hsep X_2\hsep\ldots\hsep X_m$ is a \defn{dynamic hypersequent (DHS)}. In particular, every dynamic sequent is a DHS.
\end{definition}

We can define the \defn{formula interpretation} $\iintof\cdot$ of sequents and hypersequents as follows. For a sequent $\Gamma=A_1,A_2,\ldots,A_k$, we have
$$
\intof\Gamma=\iintof{A_1,A_2,\ldots,A_k}:=A_1\OR A_2\OR\cdots\OR A_k
$$
For a dynamic sequent $X=\dsq[\ann_1]\Gamma_1\dsq[\ann_2]\Gamma_2\cdots\dsq[\ann_k]\Gamma_k$, we have
$$
\intof X:=\iintof{\dsq[\ann_1]\Gamma_1\dsq[\ann_2]\Gamma_2\cdots\dsq[\ann_n]\Gamma_n}
=\PBOX{\ann_1}\intof{\Gamma_1}\OR\PBOX{\ann_2}\intof{\Gamma_2}\OR\PBOX{\ann_n}\intof{\Gamma_n}
$$
And for a dynamic hypersequent $\hsq=X_1\hsep X_2\hsep\ldots\hsep X_m$ we have
$$
\intof\hsq=\iintof{X_1\hsep X_2\hsep\ldots\hsep X_n}
:=\BOX\intof{X_1}\OR\BOX\intof{X_2}\OR\cdots\OR\BOX\intof{X_m}
$$


\begin{figure}[!t]
  \def\myskip{\\\\[-1ex]}
  $$
  \begin{array}{c@{\hskip3em}c}
    \multicolumn{2}{c}{\mqlap{
      \vlinf{\topr}{}{\hsq\hsep X\dsq \Gamma,\top}{}
      \quad
      \vlinf{\axr}{}{\hsq\hsep X\dsq \Gamma,\cneg \atom,\atom}{}
      \quad
      \vlinf{\orr}{}{\hsq\hsep X\dsq\Gamma, A\OR B}{\hsq\hsep X\dsq\Gamma, A, B}
      \quad
      \vliinf{\andr}{}{\hsq\hsep X\dsq\Gamma, A\AND B}{
        \hsq\hsep X\dsq\Gamma, A}{
        \hsq\hsep X\dsq\Gamma, B}
    }}
    \myskip
    \multicolumn{2}{c}{\mqlap{
        \vlinf{\boxr}{}{\hsq\hsep X\dsq\Gamma, \BOX A}{
          \hsq\hsep X\dsq\Gamma \ \hsep \dsq \  A}
        \qquad
        \vlinf{\diatr}{}{\hsq\hsep X\dsq\Gamma, \DIA A}{
          \hsq\hsep X\dsq\Gamma, \DIA A, A}
        \qquad
        \vlinf{\diafr}{}{\hsq\hsep X\dsq\Gamma, \DIA A\hsep Y\dsq\Delta}{
          \hsq\hsep X\dsq\Gamma, \DIA A\hsep Y\dsq\Delta,A}
    }}
    \myskip
    \multicolumn{2}{c}{\mqlap{
      \vliiiiinf{\diapr}{}{\hsq\hsep X \dsqq{\anna\acomp\annb} \Gamma, \DIA A
        \hsep Y\dsq\Delta}{
        \hsqq{G'}\hsep Y\dsq\Delta,B_1}{
        ~\hsqq{G'}\hsep Y\dsq\Delta\dsqq{\ann\acomp B_1}B_2}{
        ~\hsqq{G'}\hsep Y\dsq\Delta\dsqq{\ann\acomp B_1\acomp B_2}B_3}{
        \cdots}{
        \hsqq{G'}\hsep Y\dsq\Delta\dsqq{\ann\acomp\annb}A}
    }}
    \myskip
    \vlinf{\pboxzr}{}{\hsq\hsep X\dsq\Gamma,\PBOX AB}{
      \hsq\hsep X\dsq\Gamma, \cneg A\dsqq{\ann\acomp A}B}
    &
    \vliinf{\pdiazr}{}{\hsq\hsep X\dsq\Gamma,\PDIA AB}{
      \hsq\hsep X\dsq\Gamma, A}{
      \hsq\hsep X\dsq\Gamma\dsqq{\ann\acomp A} B}
    \myskip
    \vlinf{\pboxr}{}{\hsq\hsep X\dsq\Gamma,\PBOX AB\dsqq{\ann\acomp A}\Delta}{
      \hsq\hsep X\dsq\Gamma, \cneg A\dsqq{\ann\acomp A}\Delta,B}
    &
    \vliinf{\pdiar}{}{\hsq\hsep X\dsq\Gamma,\PDIA AB\dsqq{\ann\acomp A}\Delta}{
      \hsq\hsep X\dsq\Gamma, A\dsqq{\ann\acomp A}\Delta}{
      \hsq\hsep X\dsq\Gamma\dsqq{\ann\acomp A}\Delta, B}
    \myskip
    \multicolumn{2}{c}{\mqlap{
      \vlinf{\atr}{}{\hsq\hsep X\dsq\Gamma\dsqq{\ann\acomp A}\Delta,\atomb}{
        \hsq\hsep X\dsq\Gamma, \atomb\dsqq{\ann\acomp A}\Delta,\atomb}
      \qquad
      \vlinf{\newr}{}{\hsq\hsep X\dsqq{\ann\acomp A}\Delta}{
        \hsq\hsep X\dsq{\!}\dsqq{\ann\acomp A}\Delta}
      \qquad
      \vlinf{\recallr}{}{\hsq\hsep X\dsq\Gamma\dsqq{\ann\acomp A}\Delta}{
        \hsq\hsep X\dsq\Gamma, \cneg A\dsqq{\ann\acomp A}\Delta}
    }}
  \end{array}
  $$
    where we have the following side conditions:\\[.4ex]
  \begin{tabular}{rl}
    in $\diapr$:~&
    $\annb$ abbreviates $B_1\cdot B_2\cdots B_n$ and $\hsq[G']$ abbreviates $\hsq\hsep X \dsqq{\anna\acomp\annb} \Gamma, \DIA A$,\\
    &
    and $\dsqq{\anna\acomp\annb'}$ does not appear in $Y$ for any non-empty prefix $\annb'$ of~$\annb$;
    \\[.15ex]
    in $\pboxzr$ and $\pdiazr$:~& $\dsq[\ann\acomp A]$ does not appear in $X$;
    \\[.15ex]
    in $\newr$:~& $\dsq$ does not appear in $X$;
    \\[.15ex]
    in $\atr$:~& $\atomb$ can be any atom, negated atom, or unit.
  \end{tabular}
  \caption{The hypersequent calculus $\HPAL$}
\label{fig:HPAL}
\end{figure}


Figure~\ref{fig:HPAL} shows the inference rules for the proof system $\HPAL$, which is the hypersequent proof system of~\cite{ler:pog:26} translated into a one-sided version, using the De Morgan dualities shown in~\eqref{eq:deMorgan}. We say that a DHS $\hsq$  is \defn{provable in $\HPAL$}, written as $\proves[\HPAL]\hsq$, if there is a derivation of $\hsq$ in $\HPAL$.

Observe that every formula $A$ is also a DHS.

\begin{lemma}\label{lem:generalised_axioms}
    Every DHS of the form $\hsq \hsep X \dsq \Gamma, A, \cneg A$ is derivable in $\HPAL$.
\end{lemma}

\begin{proof}
  By straightforward induction on $A$.
  \qed
\end{proof}

\begin{lemma}
  For every DHS $\hsq$ we have $\proves[\HPAL]\hsq$ iff\/ $\proves[\HPAL]{\intof\hsq}$.
\end{lemma}
\begin{proof}
  By routine induction on $\hsq$.
  \qed
\end{proof}

\begin{theorem}[\cite{ler:pog:26}]\label{thm:HPAL}
  $\HPAL$ is sound and complete for $\PAL$,
  i.e., $\proves[\HPAL]A$ iff\/ $\proves[\PAL]A$.
\end{theorem}

We end this section by stating that the cut rule
\begin{equation}
  \label{eq:cutr}
\vliinf{\cutr}{}{\hsq \hsep \mathcal{H} \hsep X \xMerge Y \dsq \Gamma,\Delta}{\hsq \hsep X \dsq \Gamma, A}{\mathcal{H} \hsep Y \dsq A, \Delta}  
\end{equation}
is admissible in $\HPAL$. Here, the \defn{merge} $X \xMerge Y$ of two dynamic sequents $X$ and $Y$ is defined as follows:
if $X = \dsqq{\alpha_1}\Gamma_1 \dsqq{} \cdots \dsqq{\alpha_n}\Gamma_n$ and $Y = \dsqq{\beta_1}\Delta_1 \dsqq{} \cdots \dsqq{\beta_k}\Delta_k$, then $X \xMerge Y$ is obtained by merging any two components $\dsqq{\alpha_i}\Gamma_i$ and $\dsqq{\beta_j}\Delta_j$ whenever $\alpha_i = \beta_j$, resulting in $\dsqq{\alpha_i}\Gamma_i, \Delta_j$, and simply juxtaposing the others. Consequently, if $X = X' \dsq \Gamma$ and $Y = Y' \dsq \Delta$ the resulting dynamic sequent is $X \xMerge Y = X' \xMerge Y' \dsq \Gamma,\Delta$.
For instance, if $X = \Gamma \dsq \Gamma' \dsqq{\beta} \Gamma''$ and $Y = \Delta \dsq \Delta' \dsqq{\gamma} \Delta''$ (where $\alpha,\beta,\gamma$ are distinct) then $X \xMerge Y = \Gamma,\Delta \dsq \Gamma',\Delta' \dsqq{\beta}\Gamma'' \dsqq{\gamma}\Delta''$. For more details, see~\cite[\textsection4]{ler:pog:26}.
\begin{theorem}[\cite{ler:pog:26}]\label{thm:cut_HPAL}
The $\cutr$-rule is admissible in $\HPAL$.
\end{theorem}
\begin{proof}
  For the one-sided version of the proof system, as we presented it here, the proof is exactly the same as for the two-sided version, as presented in \cite[\textsection6]{ler:pog:26}.
  \qed
\end{proof}
We will write $\proves[\HPAL+\cutr]A$ whenever $A$ is provable in  $\HPAL$ extended with the $\cutr$-rule. We then have immediately the following.
\begin{corollary}\label{thm:hpalcut-to-hpal}
  If $\;\proves[\HPAL+\cutr]A\;$ then $\;\proves[\HPAL]A\;$.
\end{corollary}

\section{A Deep Inference System for $\PAL$}
\label{sec:DI}


After having presented the hypersequent system $\HPAL$, we now present the open deduction system $\OPAL$ for public announcement logic. Here, the inference rules operate directly on the formulas (as defined in \Cref{sec:PAL}), so that we do not need to introduce another syntactic construct. The basic idea of open deduction~\cite{GGP:open} is that derivations are constructed by the same operations as formulas, together with a vertical composition via inference rules. For $\OPAL$, the inference rules are shown in \Cref{fig:OPAL}. The rule $\equiv$ uses the equivalence relation $\equiv$ on formulas, which is defined to be the smallest congruence relation obeying
\begin{equation}
  \label{eq:equiv}
  (A\OR B)\OR C\equiv A\OR(B\OR C)
  \hskip3em
  A\OR B\equiv B\OR A
  \hskip3em
  \bot \OR A\equiv A
\end{equation}
for  all formulas $A,B,C$.

\begin{definition}\label{def:derivation}
  The set of \defn{derivations} in $\OPAL$ is defined inductively together with the functions \defn{premise} and \defn{conclusion} from derivations to formulas, denoted by $\prem\cdot$ and $\concl\cdot$, respectively, where we use the notation $\odv{A}\deri{B}{}$ to denote a derivation $\deri$ with premise $\prem\deri=A$ and conclusion $\concl\deri=B$.
  \begin{itemize}
        \item If $A$ is a formula, then $\deri=\odf{A}$ is a derivation with $\prem\deri=\concl\deri=A$.
        
        \item If $\odv{A}{\derib}{B}{}$ and $\odv{C}{\deric}{D}{}$ are derivations, then so is $\deri=\odf{\odf\derib\ANDOR\odf\deric}=\odf{\odv{A}{\derib}{B}{}\ANDOR\odv{C}{\deric}{D}{}}$ where $\ANDOR\in\set{\AND,\OR}$, and $\prem\deri=A\ANDOR C$ and $\concl\deri=B\ANDOR D$.
        
        \item If $\odv{A}{\derib}{B}{}$ is a derivation, then so is $\deri=\odf{\BOXDIA\odf\derib}=\odf{\BOXDIA\odv{A}{\derib}{B}{}}$ where  $\BOXDIA\in\set{\BOX,\DIA,\PBOX C,\PDIA C}$ and $C$ is a formula, and  $\prem\deri=\BOXDIA A$ and $\concl\deri=\BOXDIA B$.
          
        \item If $\odv{A}{\derib}{B}{}$ and $\odv{C}{\deric}{D}{}$ are derivations,
          then so is
          $\deri=\odn{\odf\derib}{\rrule}{\odf\deric}{}=
          \odn{\odv{A}{\derib}{B}{}}{\rrule}{\odv{C}{\deric}{D}{}}{}$
          where $\vlinf{\rrule}{}{C}{B}$ is an instance of an inference rule, and
           $\prem\deri=A$ and $\concl\deri=D$.
    \end{itemize}
    If $\deri$ is a derivation with $\prem\deri=A$ and $\concl\deri=B$, using only rules from a proof system~$\sysS$, then we denote this as  $\odv{A}\deri{B}{\sysS}$. If $\prem\deri=\top$, then we can also write $\odr\deri{B}{\sysS}$.
\end{definition}

\begin{definition}
  A rule $\upsmash{\vlinf{\rrule}{}{B}{A}}$ is \defn{derivable} in a system $\sysS$ iff there is a derivation $\upsmash{\odv{A}{\deri}{B}{\sysS}}$.
\end{definition}


\def\myskip{-1.5ex}
\begin{figure}[!t]
  \vskip-2ex
  $$
  \begin{array}{c}
  \vlinf{\top}{}{\top\AND\top}{\top}
  \qquad\!
  \vlinf{\aiD}{}{\cneg\atom\OR \atom}{\top}
  \qquad\!
  \vlinf{\conD}{}{A}{A\OR A}
  \qquad\!
  \vlinf{\weakD}{}{A}{\bot}
  \qquad\!
  \vlinf{\sw}{}{(A\OR B) \OR(C \AND D)}{(A\OR C)\AND (B \OR D)}
  \qquad\!
  \vlinf{\equiv}{\proviso{if $A\equiv B$}}{A}{B}
  \\\\[\myskip]
  \vlinf{\necD}{}{\BOX\top}{\top}
  \qquad
  \vlinf{\kD}{}{\BOX A\OR\DIA B}{\BOX(A\OR B)}
  \qquad
  \vlinf{\kpU}{}{\BOX(A\AND B)}{\BOX A\AND \BOX B}
  \qquad
  \vlinf{\tD}{}{\DIA A}{A}
  \qquad
  \vlinf{\vU}{}{\BOX\BOX A}{\BOX A}
  \qquad
  \vlinf{\fU}{}{\BOX\DIA A}{\DIA A}
  \\\\[\myskip]
  \vlinf{\necPD}{}{\PBOX A \atomb}{\atomb} \dagger
  \qquad
  \vlinf{\kPD}{}{\PBOX AB \OR \PDIA AC}{\PBOX A(B \OR C)}
  \qquad
  \vlinf{\kpPU}{}{\PBOX A(B\AND C)}{\PBOX AB \AND \PBOX AC}
  \qquad
  \vlinf{\funPD}{}{\PBOX AB}{\PDIA AB}
  \\\\[\myskip]
  \vlinf{\diaPD}{}{\PBOX A\DIA B}{\DIA \PDIA A B}
  \qquad
  \vlinf{\boxPD}{}{\PBOX A\BOX B}{\BOX \PBOX A B}
  \qquad
  \vlinf{\redD}{}{\PDIA AB}{A\AND \PBOX AB}
  \qquad
  \vlinf{\negD}{}{\PBOX AB}{\cneg A}
  \end{array}
  $$
 $\dagger$ where $\atomb$ can be any atom, negated atom, or unit

  \caption{The open deduction system $\OPAL$ for Public Announcement Logic
  }  
    \label{fig:OPAL}
\end{figure}

\begin{figure}[!t]
  \vskip-2ex
  $$
  \begin{array}{c}
  \vlinf{\iD}{}{\cneg A\OR A}{\top}
  \qquad
  \vlinf{\medD}{}{\BOX(A \OR  B)}{\BOX A \OR \BOX B}
  \qquad
  \vlinf{\medvD}{}{\BOX(A \OR \BOX B)}{\BOX A \OR \BOX B}
  \qquad
  \vlinf{\medfD}{}{\BOX(A \OR \DIA B)}{\BOX A \OR \DIA B}
  \qquad
  \vlinf{\kfD}{}{\BOX(A\OR\DIA B)\OR\BOX C}{\BOX A\OR\BOX(C \OR B)}
  \\\\[\myskip]
  \vlinf{\medhPD}{}{\PBOX \ann(B \OR C)}{\PBOX \ann B \OR \PBOX \ann C}
  \qquad
  \vlinf{\khPD}{}{\PBOX \ann B \OR \PDIA\ann C}{\PBOX \ann (B\OR C)}
  \qquad
  \vlinf{\funhPD}{}{\PBOX \ann B}{\PDIA \ann B}
  \qquad
  \vlinf{\diahPD}{}{\PBOX \ann\DIA B}{\DIA \PDIA \ann B}
  \qquad
  \vlinf{\boxhPD}{}{\PBOX \ann\BOX B}{\BOX \PBOX \ann B}
  \end{array}
  $$  
  \caption{Additional derivable rules of $\OPAL$}
  \label{fig:OPALplus}
\end{figure}


One of the crucial properties of the $\OPAL$ proof system is that the general identity rule $\iD$, shown top-left in \Cref{fig:OPALplus}, is \emph{derivable}. This is similar to many sequent calculus proof systems. But deep inference allows also the other rules in \Cref{fig:OPALplus} to be derivable.

\begin{restatable}{lemma}{LemDerivableRules}\label{lem:derivable}
  The rules shown in \Cref{fig:OPALplus} are all derivable in $\OPAL$.
\end{restatable}

\begin{proof}
  For the rule $\iD$, this is a straightforward induction on $A$. The rules $\medD$, $\medvD$, $\medfD$, and~$\kfD$, can be derived by a simple combination of $\weakD$, $\conD$, and the modal rules. The remaining rules need an induction on~$\ann$. Details can be found in Appendix~\ref{app:DI}. \qed
\end{proof}

\begin{definition}
  A \defn{context} is a formula in which one atom- or unit-occurrence is replaced by a \defn{hole} $\conhole$. More precisely, contexts are generated by the grammar 
\begin{equation}
  \label{eq:context}
  \begin{array}{rcl}
  \Chole&\coloneqq& \conhole\mid
  A\AND\Chole\mid\Chole\AND A\mid A\OR\Chole\mid\Chole\OR A\mid\\[.3ex]
  &&
  \BOX\Chole\mid\DIA\Chole\mid\PBOX A\Chole\mid\PDIA A\Chole
  \end{array}
\end{equation}
If $\Chole$ is a context and $A$ is a formula, then $\Ccon A$ is the formula obtained from $\Chole$ by replacing the hole~$\conhole$ by~$A$.
\end{definition}
Deep inference allows to generalize this to derivations, \emph{i.e.},
if $\deri$ is a derivation with premise $A$ and conclusion $B$,
and $\Chole$ a context, then we can construct, by induction on $\Chole$, the following derivation
$$
\deri'
\quad=\quad
\odf{\Ccon{\odf \deri}}
\quad=\quad
\odf{\CCon{\odv{A}{\deri}{B}{}}}
\quad=\quad
\odv{\Ccon A}{\Ccon \deri}{\Ccon B}{}
$$
with premise $\prem{\deri'}=\Ccon A$ and conclusion $\concl{\deri'}=\Ccon B$.


\section{Soundness}
\label{sec:soundness}

For showing soundness of $\OPAL$, we have to show that (i) every inference rule of $\OPAL$, shown in~\Cref{fig:OPAL}  is sound (done in \Cref{lem:rule-sound} below), and (ii) each case in the inductive definition of derivations (Definition~\ref{def:derivation}) preserves soundness (done in \Cref{thm:opal-to-hpal} below).

\begin{restatable}{lemma}{LemRuleSound}\label{lem:rule-sound}
    For every rule ${\vlinf{\rrule}{}{B}{A}}$ of\/ $\OPAL$, we have\/ $\proves[\PAL]A\IMP B$ and $\proves[\HPAL]{\cneg A, B}$.
\end{restatable}

\begin{proof}
  For the rules $\top$, $\aiD$, $\conD$, $\weakD$, $\sw$, and $\equiv$ in the first row of \Cref{fig:OPAL}, we have that $\proves[\PAL]A\IMP B$ follows immediately from propositional reasoning. For the rules $\necD$, $\kD$, $\kpU$, $\tD$, $\vU$, $\fU$ in the second row of \Cref{fig:OPAL}, the corresponding implications are theorems of the modal logic $\Sfive$. The remaining implications can be easily shown either syntactically, using the reduction axioms in~\eqref{eq:axioms} and~\eqref{eq:reduction2}, or by semantic means. Then $\proves[\HPAL]{\cneg A\OR B}$ (and therefore also $\proves[\HPAL]{\cneg A, B}$) follows from \Cref{thm:HPAL}. We can also show $\proves[\HPAL]{\cneg A, B}$ directly, from which then $\proves[\PAL]A\IMP B$ would follow. Here is the case for the rule $\boxPD$ (where $\hderi_1$ and $\hderi_2$ exist by \Cref{lem:generalised_axioms}):\vadjust{\vskip-3ex}
$$
    \vlinf{\boxPD}{}{\PBOX A\BOX B}{\BOX \PBOX A B}
    \quad
    \qualto
    \vlderivation{
    \vlin{\pboxzr}{}{\DIA \PDIA A \cneg B, \PBOX A\BOX B}{
    \vlin{\boxr}{}{\cneg A, \DIA \PDIA A \cneg B \dsqq{A}\BOX B}{
    \vliq{\newr + \recallr}{}{\cneg A, \DIA \PDIA A \cneg B \dsqq{A} \ \hsep \ \dsqq{A} B}{
    \vlin{\diafr}{}{\cneg A, \DIA \PDIA A \cneg B \dsqq{A} \ \hsep \ \cneg A \dsqq{A}  B}{
    \vliin{\pdiar}{}{\cneg A, \DIA \PDIA A \cneg B \dsqq{A} \ \hsep \ \PDIA A \cneg B, \cneg A \dsqq{A} B}
    {\vlhtr{\hderi_1}{\cneg A, \DIA \PDIA A \cneg B \dsqq{A} \ \hsep \ A, \cneg A \dsqq{A}  B}}
    {\vlhtr{\hderi_2}{\cneg A, \DIA \PDIA A \cneg B \dsqq{A} \ \hsep \ \cneg A \dsqq{A} B, \cneg B}}
    }}}}}
    $$
  For the convenience of the reader, we show all cases in Appendix~\ref{app:soundness}.
  \qed
\end{proof}

\begin{restatable}{theorem}{ThmOpalToHpal}\label{thm:opal-to-hpal}
    For every derivation $\odv{A}{\deri}{B}{\OPAL}$, we have\/ $\proves[\PAL]A\IMP B$ and $\proves[\HPAL+\cutr]{\cneg A, B}$.
\end{restatable}

\begin{proof}
  We can show both, $\proves[\PAL]A\IMP B$ and  $\proves[\HPAL+\cutr]{\cneg A, B}$ by structural induction on the derivation~$\deri$ (\textit{cf.}~Definition~\ref{def:derivation}).
  In Appendix~\ref{app:soundness} we show the details for $\proves[\HPAL+\cutr]{\cneg A, B}$. Then $\proves[\PAL]A\IMP B$ follows by \Cref{thm:hpalcut-to-hpal} and \Cref{thm:HPAL}.
  \qed
\end{proof}

\begin{corollary}[Soundness]\label{thm:soundness}
  If $\proves[\OPAL]A$ then $\proves[\PAL]A$.
\end{corollary}
\begin{proof}
  Immediately from \Cref{thm:opal-to-hpal} since $A\EQU(\top\IMP A)$.
  \qed
\end{proof}
\begin{corollary}
  If $\proves[\OPAL]A$ then $\proves[\HPAL]A$.
\end{corollary}
\begin{proof}
  Immediately from \Cref{thm:soundness} and \Cref{thm:HPAL}, or directly, with the same construction as in the proof of \Cref{thm:opal-to-hpal}. 
  \qed
\end{proof}


\section{Completeness}
\label{sec:completeness}

To simplify the presentation of the completeness proof for $\OPAL$, we first show some auxiliary lemmas about the properties of $\OPAL$.

\begin{lemma}\label{lem:and}
  For any formulas $A$ and $B$, if $\;\proves[\OPAL]A\AND B\;$ then $\;\proves[\OPAL]A\;$ and $\;\proves[\OPAL]B\;$. 
\end{lemma}

\begin{proof}
  Observe that the only rules that can have an $\AND$ as outermost connective in the conclusion are $\top$ and~$\equiv$. In order to deal with~$\equiv$ we define the notions of a $\bot$-formula $\Abot$ and $\bot$-context $\Cbhole$ as follows:
  $$
  \begin{array}{rcl}
    \Abot&\coloneq&\bot\mid\Abot\OR\bot\mid\bot\OR\Abot\\
    \Cbhole&\coloneq&\conhole\mid\Abot\OR\Cbhole\mid\Cbhole\OR\Abot
  \end{array}
  $$
  i.e., for every $\bot$-formula $\Abot$, we have $\Abot\equiv\bot$, and for every $\bot$-context $\Cbhole$ and formula $B$, we have $\Cbcon B\equiv B$.
  We are now showing a more general statement: for any $\bot$-context $\Cbhole$, if $\proves[\OPAL]{\Cbcon{A\AND B}}$ then $\proves[\OPAL]{\Cbcon A}$ and $\proves[\OPAL]{\Cbcon B}$. We proceed by induction on the derivation $\odr{\deri}{\Cbcon{A\AND B}}{\OPAL}$. We have the following cases how this derivation can be obtained:
  $$
  \scalebox{1}{$
  (1)\colon\odr{\deri'}{\Cbcon{\odn{\top}{\top}{\top\AND\top}{}}}{\OPAL}
  \qquad
  (2)\colon\odr{\deri'}{\Cbcon{\odv{A'}{\deri_A}{A}{}\AND\odv{B'}{\deri_B}{B}{}}}{\OPAL}
  \qquad
  (3)\colon\odr{\deri'}{\odn{\Cbpcon{A'\AND B'}}{\equiv}{\Cbcon{A\AND B}}{}}{\OPAL}
  $}
  $$
  In Case~(1), we have $A=B=\top$, and we are done immediately, as $\deri'$ is our desired derivation. In Cases (2) and~(3), we apply the induction hypothesis to $\deri'$. Note that Case~(3) subsumes the cases where instead of $\equiv$, we use one of the rules $\weakD$, $\conD$, or $\sw$, as in each case we have that premise and conclusion are equivalent modulo $\equiv$.
  \qed
\end{proof}

\begin{lemma}\label{lem:box}
  For any formula $A$, if $\;\proves[\OPAL]\BOX A\;$ then also $\;\proves[\OPAL]A\;$. 
\end{lemma}

\begin{proof}
  The only rules that have a $\BOX$ as outermost connective in the conclusion are $\necD$, $\kpU$, $\vU$ and $\fU$. We proceed by induction on the derivation $\odr{\deri}{\BOX A}{\OPAL}$ and the formula $A$, and we have the following cases:
  $$
  1\colon
  \odn{\top}{\necD}{\BOX\top}{}
  \quad
  2\colon
  \odr{\deri'}{\odn{\BOX A'}{\vU}{\BOX\BOX A'}{}}{\OPAL}
  \quad
  3\colon
  \odr{\deri'}{\odn{\DIA A'}{\fU}{\BOX\DIA A'}{}}{\OPAL}
  \quad
  4\colon
  \odr{\deri'}{\BOX \odv{A'}{\deri_A}{A}{}}{\OPAL}
  \quad
  5\colon
  \odr{\deri'}{\odn{\BOX A'\AND\BOX A''}{\kpU}{\BOX(A'\AND A'')}{}}{\OPAL}
  $$
  In the first three cases we are done immediately. In Case~4, we proceed by induction hypothesis (as in the previous lemma), and in Case~5, we apply \Cref{lem:and}, to get $\proves[\OPAL]{\BOX A'}$ and $\proves[\OPAL]\BOX A''$, to which we apply the induction hypothesis, to obtain $\proves[\OPAL]A'$ and $\proves[\OPAL]A''$, from which we get immediately $\proves[\OPAL]{A'\AND A''}$.
  \qed
\end{proof}

\begin{definition}
  A \defn{tame context} is a context in which the hole $\conhole$ does not occur inside an $\AND$, nor in the scope of a $\DIA$ or $\PDIA A$ modality, \textit{i.e.}, a tame context is generated by the grammar 
\begin{equation}
  \label{eq:tamecontext}
  \Chole\coloneqq
  \conhole\mid
  A\OR\Chole\mid\Chole\OR A\mid
  \BOX\Chole\mid\PBOX A\Chole
\end{equation}
\end{definition}

\begin{lemma}\label{lem:and-2}
  Let $\Chole$ be a tame context and let $A$ and $B$ be formulas.
  Then there is a derivation
  $$\odv{\Ccon{A}\AND \Ccon{B}}{\deri}{\Ccon{A\AND B}}{\OPAL}$$
\end{lemma}

\begin{proof}
  We construct $\deri$ by induction on $\Chole$. The case of $\Chole=\conhole$ is trivial. The cases for $\Chole=D\OR\Cphole$ and $\Chole=\BOX\Cphole$ and $\Chole=\PBOX D\Cphole$ are as follows (respectively):
  $$
  \scalebox{.9}{$
  \odn{(D\OR\Cpcon A)\AND(D\OR\Cpcon B)}{\sw}{
    \odn{D\OR D}{\conD}{D}{}\OR
    \odv{\Cpcon A\AND\Cpcon B}{\deri'}{\Cpcon{A\AND B}}{\OPAL}
  }{}
  \quad  
  \odn{\BOX\Cpcon A\AND\BOX\Cpcon B}{\kboxU}{
    \BOXof{\odv{\Cpcon A\AND\Cpcon B}{\deri'}{\Cpcon{A\AND B}}{\OPAL}}
  }{}
  \quad
  \odn{\PBOX D\Cpcon A\AND\PBOX D\Cpcon B}{\kpboxU}{
    \PBOXof D{\odv{\Cpcon A\AND\Cpcon B}{\deri'}{\Cpcon{A\AND B}}{\OPAL}}
  }{}
  $}
  $$
  where $\deri'$ exists by induction hypothesis, and the case  $\Chole=\Cphole\OR D$ is similar to $\Chole=D\OR\Cphole$.
  \qed
\end{proof}

\begin{lemma}\label{lem:and-n}
  Let $\Chole$ be a tame context and let $A_1,\ldots,A_n$ be formulas for $n\ge0$.
  Then there is a derivation
  $$\odv{\Ccon{A_1}\AND\cdots\AND \Ccon{A_n}}{\deri}{\Ccon{A_1\AND\cdots\AND A_n}}{\OPAL}$$
\end{lemma}

\begin{proof}
  We proceed by induction on $n$. The base case is $n=0$. We can show by induction on $\Chole$ that there is a derivation from $\top$ to $\Ccon{\top}$ via:
  $$
  \scalebox{.9}{$
  \odn{\top}{\equiv}{
    \odn{\bot}{\weakD}D{}\OR
    \odv{\top}{\deri'}{\Cpcon{\top}}{\OPAL}
  }{}
  \qquad  
  \odn{\top}{\eboxD}{
    \BOX{\odv{\top}{\deri'}{\Cpcon{\top}}{\OPAL}}
  }{}
  \qquad
  \odn{\top}{\epboxD}{
    \PBOX D{\odv{\top}{\deri'}{\Cpcon{\top}}{\OPAL}}
  }{}
  $}
  $$
  and the inductive case follows immediately from \Cref{lem:and-2}.
  \qed
\end{proof}

\begin{restatable}{lemma}{LemHpalOpal}\label{lem:hpal-opal}
  For every $\HPAL$ inference rule
  \begin{equation}
    \label{eq:hpalrule}
    \vliiinf{\hpalr}{}{\hsq}{\hsq_1}{\cdots}{\hsq_n}
  \end{equation}
  with $n\ge0$, there is an $\OPAL$ derivation
  $$
  \odv{\intof{\hsq_1}\AND\cdots\AND\intof{\hsq_n}}{\deri}{\intof\hsq}{\OPAL}
  $$
\end{restatable}

\begin{proof}
 We construct $\deri$
 for each rule of $\HPAL$. Note that for the axioms $\topr$ and $\axr$, as well as for the branching rules $\andr$, $\diapr$, $\pdiazr$, and $\pdiar$, we make crucial use of \Cref{lem:and-2} and \Cref{lem:and-n}. The most involved cases are the rules $\diapr$ and $\recallr$, for which we show the derivation $\deri$ in \Cref{fig:hpal-opal}. 
For the convenience of the reader, a complete list is shown in Appendix~\ref{app:completeness}.
  \qed
\end{proof}


\begin{figure}[!t]
  \scalebox{.9}{$\vliiiiinf{\diapr}{}{\hsq\hsep X \dsqq{\anna\acomp\annb} \Gamma, \DIA A
        \hsep Y\dsq\Delta}{
        \hsqq{G'}\hsep Y\dsq\Delta,B_1}{
        ~\hsqq{G'}\hsep Y\dsq\Delta\dsqq{\ann\acomp B_1}B_2}{
        ~\hsqq{G'}\hsep Y\dsq\Delta\dsqq{\ann\acomp B_1\acomp B_2}B_3}{
        \cdots}{
        \hsqq{G'}\hsep Y\dsq\Delta\dsqq{\ann\acomp\annb}A}
    \quad\leadsto $}\\[1ex]
  \scalebox{.73}{%
    $
    \odn{
      \odv{G\OR\BOXof{\intof Y\OR\PBOXof\ann{\intof\Delta\OR B_1}})
        \AND
        \left(G\OR\BOXof{\intof Y\OR
          \odn{\PBOX\ann\intof\Delta\OR\PBOX{\ann.B_1}B_2}{
            \medhPD}{
            \PBOXof\ann{\intof\Delta\OR \PBOX{B_1}B_2}}{}}\right)
        \AND\cdots\AND
        \left(G\OR\BOXof{\intof Y\OR
          \odn{\PBOX\ann\intof\Delta\OR\PBOX{\ann \cdot \annb}A}{
            \medhPD}{
            \PBOXof\ann{\intof\Delta\OR \PBOX{\annb}A}}{}}\right)
      }{
        \deri_1}{
        G\OR\BOXof{\intof Y\OR\PBOXof\ann{\intof\Delta\OR
            \odv{B_1\AND\PBOX{B_1}B_2\AND\PBOX{B_1.B_2}B_3
              \AND\cdots\AND\PBOX{\annb}A}{\deri_0}{
              \PDIA\annb A}{}}}}{}}{
      \equiv}{
        \intof\hsq\OR
        \odn{\BOX(\intof X\OR\PBOX{\ann \cdot \annb}(\intof\Gamma\OR\DIA A))
          \OR
          \BOXof{\intof Y\OR\odn{\PBOX{\ann}(\intof \Delta\OR\PDIA{\annb} A)}{
              \khPD}{\PBOX{\ann}\intof\Delta\OR\PDIA{\ann \cdot \annb} A}{}}
    }{\kfD}{
      \BOXof{
        \intof X\OR
        \odn{\PBOX{\ann \cdot \annb}(\intof \Gamma\OR\DIA A)\OR
        \odn{\DIA\PDIA{\ann \cdot \annb} A}{\diahPD}{\PBOX{\ann \cdot \annb}\DIA A}{}}{
          \medhPD}{
        \PBOXof{\ann \cdot \annb}{\intof\Gamma\OR\odn{\DIA A\OR\DIA A}{\conD}{\DIA A}{}}}{}}
      \OR\BOX(\intof Y\OR\PBOX{\ann}\intof\Delta)
        }{}}{}
    $
    }
where
    $G=\intof{(\hsq')}=\intof{(\hsq\hsep X \dsqq{\anna\acomp\annb} \Gamma, \DIA A)}
    =\intof\hsq\OR\BOX(\intof X\OR\PBOX{\ann\acomp\annb}(\intof\Gamma\OR\DIA A))$,
    and $\deri_1$ is obtained from \Cref{lem:and-n}, and $\deri_0$ is 
  \scalebox{.9}{$
    \od{\odI{\odI{\odh{B_1\AND\PBOX{B_1}B_2\AND\PBOX{B_1 \cdot B_2}B_3
            \AND\cdots\AND\PBOX{B_1 \cdot B_2 \cdots B_n}A}}{
          (n-1)\times\medPD}{
        B_1\AND\PBOX{B_1}(B_2\AND\PBOX{B_2}(B_3
        \AND\cdots\PBOX{B_{n-1}}(B_n\AND\PBOX{B_n}A)\cdots))}{}}{
        (n-1)\times\redD}{
        \PDIA{B_1 \cdot B_2 \cdots B_n}A}{}}
    $}
  \\[3ex]
  \scalebox{.9}{$
      \vlinf{\recallr}{}{\hsq\hsep X\dsq\Gamma\dsqq{\ann\acomp A}\Delta}{
        \hsq\hsep X\dsq\Gamma, \cneg A\dsqq{\ann\acomp A}\Delta}
   \qualto
    \intof\hsq\OR \BOXof{\intof X \OR
      \odn{\odn{\PBOX\ann(\intof\Gamma\OR\cneg A)}{
          \khPD}{
          \PBOX\ann{\intof\Gamma}\OR\PDIA \ann \cneg A}{}
        \OR \PBOX {\ann\acomp A}\intof\Delta}{
        \equiv}{
        \PBOX\ann{\intof\Gamma}\OR
        \odn{\odn{\PDIA \ann \cneg A}{\funhPD}{\PBOX \ann \cneg A}{}\OR \PBOX {\ann\acomp A}\intof\Delta}{
          \medhPD}{
          \PBOXof \ann {
            \odn{\odn{\cneg A}{\negD}{\PBOX{A}\intof\Delta}{}
              \OR \PBOX{A}\intof\Delta }{
              \conD}{
              \PBOX{A}\intof\Delta 
            }{}}}{}}{}}
    $}
  \caption{$\OPAL$-derivations for the $\HPAL$ rules $\diapr$ and $\recallr$}
  \label{fig:hpal-opal}
\end{figure}
 
\begin{theorem}\label{thm:hpal-to-opal}
  If $\;\proves[\HPAL]A\;$ then $\;\proves[\OPAL]A\;$.
\end{theorem}

\begin{proof}
  By induction on the $\HPAL$ derivation of the hypersequent $\hsq=A$, we can, by applying \Cref{lem:hpal-opal} at each step, construct an $\OPAL$-derivation with conclusion $\intof\hsq=\BOX A$. By \Cref{lem:box}, this yields $\;\proves[\OPAL]A\;$.
  \qed
\end{proof}

\begin{corollary}[Completeness]
  If $\;\proves[\PAL]A\;$ then $\;\proves[\OPAL]A\;$.
\end{corollary}

\begin{proof}
  Let $A$ be a formula with $\proves[\PAL]A$. By \Cref{thm:HPAL}, we have $\proves[\HPAL]A$. Then, by \Cref{thm:hpal-to-opal}, we get $\proves[\OPAL]A$.
  \qed
\end{proof}


\section{Cut Elimination}\label{sec:cut}

The general cut rule in deep inference systems is
\begin{equation}
  \label{eq:cut}
  \vlinf{\iU}{}{\bot}{A\AND\cneg A}
\end{equation}
It is dual to the $\iD$-rule shown in \Cref{fig:OPALplus}, and therefore, soundness of the $\iU$-rule follows immediately, and we can extend \Cref{thm:opal-to-hpal} as follows:

\begin{theorem}\label{thm:opalcut-to-hpalcut}
  If $\;\proves[\OPAL+\iU]A\;$ then $\;\proves[\HPAL+\cutr]A\;$.
\end{theorem}

This is enough to prove cut admissibility for $\OPAL$:

\begin{corollary}
  If $\proves[\OPAL+\iU]A$ then $\proves[\OPAL]A$.
\end{corollary}

\begin{proof}
  Immediately from \Cref{thm:opalcut-to-hpalcut}, \Cref{thm:hpalcut-to-hpal}, and \Cref{thm:hpal-to-opal}.
  \qed
\end{proof}

\begin{remark}
Here, we have used the cut elimination theorem of the hypersequent system $\HPAL$ (proved in~\cite{ler:pog:26}) to prove cut elimination for $\OPAL$. It is clear that the next step should be to prove this internally, without the detour via $\HPAL$, using the methods from \cite{brunnler:06:herbrand} and \cite{SIS-IV,SIS-V}. In a deep inference system, cut elimination usually means that all \emph{up-rules} can be eliminated, which are rules that are dual to the \emph{down-rules}, that are needed for completeness.
Unfortunately, at the moment, there is not even an internal cut elimination proof for a deep inference system for $\Sfive$. To our knowledge, there are two cut-free systems for $\Sfive$, the one we presented here and the one by Stouppa~\cite{stouppa}. Even though, they both use essentially the same hypersequent system to show cut-free completeness, they are very different from each other. Their difference in the propositional part is merely cosmetic\footnote{For example, they use different forms of the $\sw$-rule and the formula equivalence $\equiv$, but these differences have been discussed elsewhere (see \textit{e.g.} \cite{strassburger2026interpolation}).}, but their difference in the modal rules is more fundamental. Stouppa's system does not need the $\kpU$-rule, which would be considered an up-rule in the system for $\Sfour$, presented in~\cite{strassburger2026interpolation}. Furthermore, Stouppa's system uses the rules
\begin{equation}
  \label{eq:t45}
  \vlinf{\tU}{}{A}{\BOX A}
  \qquad\qquad
  \vlinf{\vD}{}{\DIA A}{\DIA \DIA A}
  \qquad\qquad
  \vlinf{\fD}{}{\BOX A}{\DIA \BOX A}
\end{equation}
which would be considered the up-rules for our system, whereas our rules $\tD$, $\vU$, and $\fU$ would be the up-rules for Stouppa's system. To make things even more confusing, in the system for $\Sfour$, presented in~\cite{strassburger2026interpolation}, the rules $\tD$ and $\vD$ are the down-rules, and the rules $\tU$ and~$\vU$ are the up-rules. For this reason, we did not in this paper use the \hbox{usual $\downarrow$- and $\uparrow$-symbols} in the rule names, as it is common in deep inference~\cite{gug:str:01,brunnler:tiu:01}. We have to leave it to future research to unravel the mystery of \emph{up} and \emph{down} for $\Sfive$.
\end{remark}


\newcommand{\msubsection}[1]{\medskip\noindent{\bf #1:}\;}
\section{Conclusion}\label{sec:conclusion}

In this paper we presented an open deduction system for $\PAL$, which is the first deep inference system for a dynamic epistemic logic. We established soundness and completeness in a way that allows the reader to compare it to the recently introduced hypersequent calculus~\cite{ler:pog:26}.
Our work can be seen as the starting point for the investigation of the proof theory of $\DEL$ through the route of deep inference, as it opens up many research questions, some of which we briefly discuss below.

\msubsection{Multi-agent systems}
The first obvious extension of our system would be to allow for more than one epistemic modality pair, \textit{i.e.} we have a finite set $\agents=\set{\agenta,\agentb,\ldots}$ of \defn{agents}, and each agent has its own pair of modalities $\BOX_a$ and $\DIA_a$. The inference rules are very easy to adjust (simply replace $\BOX$ and $\DIA$ in each rule in~\Cref{fig:OPAL} by $\BOX_a$ and $\DIA_a$, respectively). For this, deep inference is better suited than hypersequents. However, the completeness proof for the multi-agent system would be a bit more involved, as there is no hypersequent system to which we can refer. 

\msubsection{General epistemic actions}
Another natural extension would be to consider not only public announcement but any possible epistemic action, and to provide an open deduction system for the full logic of epistemic action models \cite{baltaga1998_PAL_and_DEL,DEL}. In this logic, communication is represented by \emph{action models} $ \actionM = \tuple{ \Evt, \sim, \pre}$, where $\Evt=\set{\action, \actionb,\ldots}$ is a finite set of \emph{epistemic actions} that may be indistinguishable from the agents' perspective, written as $\action \sim \actionb$, and that can be executed only when their precondition is satisfied, \textit{i.e.} when we have $\forcesMw \pre(\action)$. This logic generalises $\PAL$, as a public announcement $\PBOX A$ can be represented by a model of only a single action $\set \action$ with $\pre(\action) = A$. The reduction axioms are similar to those of $\PAL$ but the one for the epistemic modality now becomes $\PBOX {\action}\BOX B \EQU (\pre(\action) \IMP \bigwedge_{\action \sim \actionb} \BOX \PBOX {\actionb}B)$.
It would be interesting to see how to extend $\OPAL$ for dealing with such more general epistemic actions.
We could adopt the same methodology and show soundness, completeness and cut-elimination via the dynamic hypersequent calculus of \cite{ler:pog:DEL} that extends the one for $\PAL$ \cite{ler:pog:26}.

\msubsection{Proof complexity}
In this paper we translated between the hypersequent system $\HPAL$ and the open deduction system $\OPAL$, and in the introduction we briefly mentioned other formalisms like tableaux, and labelled and non-labelled sequent calculi. An interesting question is how all these proof systems for $\PAL$ behave with each other with respect to the size of their proofs.

\msubsection{More deep inference}
Normally, deep inference allows to apply inference rules anywhere inside a formula context. However, in Definition~\ref{def:derivation}, we do not allow inferences inside the announcement formulas. The reason is that the validity of $A\IMP B$ does not imply the validity of $\PBOX{A}{C}\IMP\PBOX BC$. However, if $A\EQU B$, then also $\PBOX{A}{C}\EQU\PBOX BC$. This means that we could add the inference rule $\;\vlinf{\annD}{}{\PBOX{B}{C}}{\PBOX{A}{C}}\;$ with the proviso that we have $A\EQU B$. This raises the question if this could lead to shorter proofs.

\msubsection{Intuitionistic PAL} 
Although the underlying logic of dynamic epistemic logics is usually classical logic, we may also consider systems that employ intuitionistic logic as a basis, to represent a different kind of knowledge. There are several proposals in the literature for what Intuitionistic Public Announcement Logic ($\IPAL$) could be. First, there is the work \cite{ma2014:IPAL}, which extends the intuitionistic logic $\IK$ \cite{servi1984:IK,simpson1994:IK} and the logic $\MIPC$ \cite{prior:time,bull:mipc} with dynamic modalities for announcements.
  Apart from the Hilbert-sytle axiomatisation, that extends the one of $\IK$ (resp. of $\MIPC$) with the same reduction axioms as for classical $\PAL$, the only calculi that exist for this logic are the labelled sequent calculi of \cite{balbiani2016:IPAL} and \cite{nomura:IPAL}, and the non-labelled sequent calculus of \cite{lin:Gao:Zhao:IPAL}.
  Second, the work in \cite{bal:dit:gen:ler:IPAL} proposes a generalisation of $\IPAL$ with a binary modality, with only a Hilbert-style axiomatisation. And third, there is the work \cite{pavlova2022:IPAL}, which extends intuitionistic epistemic logic $\IEL$ by \cite{artemov} with public announcement modalities. For each of them, deep inference could provide simple proof systems, that do not need labels.

\msubsection{Interpolation}
Since $\PAL$ is equivalent to (multi-modal) $\Sfive$, it is known to satisfy Craig interpolation (if $A \IMP B$ is a theorem then there is $C$ in the common vocabulary such that both $A \IMP C$ and $C \IMP B$ are also theorems). In fact, the syntactic proof of  \cite{liu2026sequent} for interpolation for $\PAL$ uses an  adaption of Maehara's method \cite{maehara:CIP} to a non-labelled sequent calculus, where dynamic rules simulate reduction axioms, i.e., they translate $\PAL$-formulas into announcement-free $\Sfive$-formulas during proof search. Consequently, the interpolant is an announcement-free formula.
However, if an interpolant $C$ of $A \IMP B$ provides an explanation of the truth of $A \IMP B$, then one would expect such an explanation to be expressed in the same language as $A \IMP B$. In particular, announcements $\PBOX \ann$ should appear in the interpolant if $\PBOX \ann$ occurs in both $A$ and $B$.
But the interpolants computed in \cite{liu2026sequent} do not display any announcement.
Moreover, since the announcement-free translation of a $\PAL$-formulas can be exponentially longer~\cite{DEL}, the interpolant thereby obtained might be incredibly long, and translating it back into a $\PAL$-formula  might be computationally complex.
Defining an algorithm that can compute the interpolant \emph{directly in the language of $\PAL$} is therefore valuable in itself. Since $\OPAL$ is cut-free, it naturally appears as a promising calculus for defining such an interpolation algorithm.

\subsubsection{AI Disclaimer} This paper has been written by humans for humans. No AI tool has been used in the research or writing process.

%
%
\bibliographystyle{splncs04}
\bibliography{pal-di-refs}

\clearpage

\appendix

    \def\myadjust{\vadjust{\vskip-5ex}}
    \def\myadjustb{\vadjust{\vskip-3ex}}
    \def\myadjustc{\vadjust{\vskip-1ex}}

\section{Missing proofs for \Cref{sec:DI}}\label{app:DI}

\LemDerivableRules*

\begin{proof}
  For the identity rule $\iD$ we proceed by induction on $A$. When $A=\atom$ is an atom, we simply use the atomic identity rule $\aiD$. The other cases are handled with the following derivations. \myadjustc
      $$
        \odn{{\odn{\top}{\iD}{\cneg B \OR B}{} \AND \odn{\top}{\iD}{\cneg C \OR C}{}}{}}{\switchD}{(\cneg B \OR \cneg C) \OR (B \AND C)}{}
    \qquad
        \odn{\top}{\necD}{
        \odn{\BOX \odn{\top}{\iD}{\cneg B \OR B}{}}
        {\kD}{\DIA \cneg B \OR \BOX B}{}}{}
    \qquad
        \odn{\top}{\necPD}{\odn{\PBOX B \odn{ \top}{\iD}{\cneg C \OR C}{}}
        {\kPalD}{\PDIA B \cneg C \OR \PBOX BC}{}}{}
    $$
  For $\medD,\medvD,\medfD,\kfD$ and $\medhPD$, we use the following derivations.
$$
  \odn{\BOX \odn{A}{\equiv}{A \OR \odn{\bot}{\weakD}{B}{}}{} \OR \BOX \odn{B}{\equiv}{\odn{\bot}{\weakD}{A}{} \OR B}{}}{\conD}{\BOX(A \OR B)}{}
\qquad
    \odn{\BOX B\OR\odn{\BOX A}{\vU}{\BOX\BOX A}{}}{\medD}{\BOX(B\OR\BOX A)}{}
\qquad
    \odn{\BOX B\OR\odn{\DIA A}{\fU}{\BOX\DIA A}{}}{\medD}{\BOX(B\OR\DIA A)}{}
$$
$$
    \odn{\BOX A\OR
    \odn{\BOX(C\OR B)}{\kD}{\BOX C \OR \DIA B}{}
    }{\equiv}{
    \odn{\BOX A \OR\DIA B}{\medfD}{\BOX(A \OR\DIA B)}{}
    \OR\BOX C}{}
\qquad
    \odn{
    \PBOX \ann \odn{B}{\equiv}{B \OR \odn{\bot}{\weakD}{C}{}}{}
    \OR
    \PBOX \ann \odn{C}{\equiv}{\odn{\bot}{\weakD}{B}{} \OR C}{}
    }{\conD}{\PBOX \ann (B \OR C)}{}
$$  

Finally, for $\khPD,\funhPD,\diahPD$ and $\boxhPD$, we proceed by induction on $\ann$. The case where $\ann$ is empty is trivial. For the inductive case, where $\ann = \ann' \cdot A$, we use the induction hypothesis and the corresponding rules $\kPD,\funPD,\diaPD$ and $\boxPD$. This is straightforward.
\qed
\end{proof}


\section{Missing proofs for \Cref{sec:soundness}}\label{app:soundness}

For proving that $\OPAL$ is sound with respect to $\HPAL$, we need to show that whenever there is a derivation $\deri$ from $A$ to $B$ in $\OPAL$ then there is a proof $\hderi$ of $A \IMP B$ in $\HPAL$.
To do so, we first show that all inference rules of $\OPAL$ are sound.

\LemRuleSound*

\begin{proof}
    We show that, for each rule $\vlinf{\rrule}{}{B}{A}$ in $\OPAL$ there is an $\HPAL$ derivation of $\cneg A, B$. We consider here the dynamic rules (as for propositional and modal rules, this is straightforward). We show $\proves[\HPAL]{\cneg A, B}$ by means of the following derivations, where every $\hderi_i$ exists by \Cref{lem:generalised_axioms}.

\begin{itemize}
    \item For $\necPD$ we have the three following derivations, depending on whether $atom$ is a (negated) atom or unit:
    $$\vlinf{\necPD}{}{\PBOX A \atom}{\atom}
    \qualto
   \vlderivation{
    \vlin{\pboxzr}{}{\cneg \atom, \PBOX A \atom}{
    \vlin{\atr}{}{ \cneg \atom, \cneg A \dsqq{A} \atom }{\vlin{\axr}{}{\atom,\cneg \atom, \cneg A \dsqq{A} \atom}{}}
    }}
    \qquad
    \vlderivation{
    \vlin{\pboxzr}{}{\bot, \PBOX A \top}{
    \vlin{\atr}{}{ \bot, \cneg A \dsqq{A} \top }{\vlin{\topr}{}{\top,\bot, \cneg A \dsqq{A} \top}{}}
    }}
    \qquad
    \vlinf{\topr}{}{\top, \PBOX A \bot}{}
    $$
    \item For $\kPD$ we have:\myadjust
    $$
    \vlinf{\kPD}{}{\PBOX AB \OR \PDIA AC}{\PBOX A(B \OR C)}
    \qualto
    {\small
    \vlderivation{
    \vlin{\orr}{}{\PDIA A(\cneg B \AND \cneg C), \PBOX AB \OR \PDIA AC}{
    \vlin{\pboxzr}{}{\PDIA A(\cneg B \AND \cneg C),\PBOX AB, \PDIA AC}{
    \vliin{\pdiar}{}{\cneg A, \PDIA A(\cneg B \AND \cneg C), \PDIA AC \dsqq{A} B}
    {\vlhtr{\hderi_1}{A,\cneg A,\PDIA A(\cneg B \AND \cneg C) \dsqq{A} B }}
    {\vliin{\pdiar}{}{\cneg A, \PDIA A(\cneg B \AND \cneg C) \dsqq{A} B,C}
    {\vlhtr{\hderi_2}{A,\cneg A \dsqq{A} B,C}}
    {\vliin{\andr}{}{\cneg A \dsqq{A} B, C, \cneg B \AND \cneg C}
    {\vlhtr{\hderi_3}{\cneg A \dsqq{A} B, C, \cneg B}}
    {\vlhtr{\hderi_4}{\cneg A \dsqq{A} B, C, \cneg C}}}}
    }}}}
    $$
    \item For $\kpPU$ we have:\myadjust
    $$
    \vlinf{\kpPU}{}{\PBOX A(B\AND C)}{\PBOX AB \AND \PBOX AC}
    \qualto
    {\small\vlderivation{
    \vlin{\orr}{}{\PDIA A \cneg B \OR \PDIA A \cneg C, \PBOX A(B\AND C)}{
    \vlin{\pboxzr}{}{\PDIA A \cneg B, \PDIA A \cneg C,\PBOX A(B\AND C)}{
    \vliin{\pdiar}{}{\cneg A, \PDIA A \cneg B, \PDIA A \cneg C \dsqq{A} B \AND C}
    {\vlhtr{\hderi_1}{A,\cneg A, \PDIA A \cneg C \dsqq{A} B \AND C}}
    {\vliin{\pdiar}{}{\cneg A, \PDIA A \cneg C \dsqq{A} B \AND C, \cneg B}
    {\vlhtr{\hderi_2}{A,\cneg A \dsqq{A} B \AND C, \cneg B}}
    {\vliin{\andr}{}{\cneg A \dsqq{A} B \AND C, \cneg B, \cneg C}
    {\vlhtr{\hderi_3}{\cneg A \dsqq{A} B, \cneg B, \cneg C}}
    {\vlhtr{\hderi_4}{\cneg A \dsqq{A} C, \cneg B, \cneg C}}}}
    }}}}
    $$
    \item For $\funPD$ we have:\myadjustb
    $$
    \vlinf{\funPD}{}{\PBOX AB}{\PDIA AB}
    \qualto
    \vlderivation{
    \vlin{\pboxzr}{}{\PBOX A \cneg B,\PBOX AB}{
    \vlin{\pboxr}{}{\cneg A, \PBOX AB \dsqq{A} \cneg B}{
    \vlhtr{\hderi_1}{\cneg A, \cneg A \dsqq{A} B, \cneg B}
    }}}
    $$
    \pagebreak
    \item For $\diaPD$ we have: \myadjust
    $$
    \vlinf{\diaPD}{}{\PBOX A\DIA B}{\DIA \PDIA A B}
    \qualto
    \vlderivation{
    \vlin{\pboxzr}{}{\BOX \PBOX A \cneg B, \PBOX A\DIA B}{
    \vlin{\boxr}{}{\cneg A, \BOX \PBOX A\cneg B \dsqq{A} \DIA B}{
    \vlin{\pboxzr}{}{\cneg A \dsqq{A} \DIA B \hsep \PBOX A\cneg B}{
    \vlin{\diafr}{}{\cneg A \dsqq{A} \DIA B \hsep \cneg A \dsqq{A} \cneg B}{
    \vlhtr{\hderi_1}{\cneg A \dsqq{A} \DIA B \hsep \cneg A \dsqq{A} B, \cneg B}
    }}}}}
    $$
    \item For $\boxPD$ we have:\myadjust
    $$
    \vlinf{\boxPD}{}{\PBOX A\BOX B}{\BOX \PBOX A B}
    \qualto
    \vlderivation{
    \vlin{\pboxzr}{}{\DIA \PDIA A \cneg B, \PBOX A\BOX B}{
    \vlin{\boxr}{}{\cneg A, \DIA \PDIA A \cneg B,\dsqq{A}\BOX B}{
    \vliq{\newr + \recallr}{}{\cneg A, \DIA \PDIA A \cneg B,\dsqq{A} \ \hsep \dsqq{A} B}{
    \vlin{\diafr}{}{\cneg A, \DIA \PDIA A \cneg B,\dsqq{A} \ \hsep \cneg A \dsqq{A}  B}{
    \vliin{\pdiar}{}{\cneg A, \DIA \PDIA A \cneg B,\dsqq{A} \ \hsep \PDIA A \cneg B, \cneg A \dsqq{A} B}
    {\vlhtr{\hderi_1}{\cneg A, \DIA \PDIA A \cneg B,\dsqq{A} \ \hsep A, \cneg A \dsqq{A}  B}}
    {\vlhtr{\hderi_2}{\cneg A, \DIA \PDIA A \cneg B,\dsqq{A} \ \hsep \cneg A \dsqq{A} B, \cneg B}}
    }}}}}
    $$
    \item For $\redD$ we have:\myadjust
    $$
    \vlinf{\redD}{}{\PDIA AB}{A\AND \PBOX AB}
    \qualto
    \vlderivation{
    \vlin{\orr}{}{\cneg A \OR \PDIA A\cneg B, \PDIA AB}{
    \vliin{\pdiazr}{}{\cneg A, \PDIA A\cneg B, \PDIA AB}
    {\vlhtr{\hderi_1}{A, \cneg A, \PDIA AB}}
    {\vliin{\pdiazr}{}{\cneg A, \PDIA AB \dsqq{A} \cneg B}
    {\vlhtr{\hderi_2}{A, \cneg A \dsqq{A} \cneg B}}
    {\vlhtr{\hderi_3}{ \cneg A \dsqq{A} \cneg B, B }}}
    } }
    $$

    \item For $\negD$ we have:\myadjustb
    $$
    \vlinf{\negD}{}{\PBOX AB}{\cneg A}
    \qualto
    \vlderivation{
    \vlin{\pboxzr}{}{A, \PBOX AB}{
    \vlhtr{\hderi_1}{A,\cneg A \dsqq{A} B}
    }}$$
\end{itemize}
\qed
\end{proof}

We now need to prove that each possible way of composing derivations in $\OPAL$ (see Definition~\ref{def:derivation}) preserves soundness (\Cref{thm:opal-to-hpal}). This requires the use of several rules shown to be admissible in $\HPAL$. We therefore first state some result about $\HPAL$ before proceeding.

\begin{lemma}[\cite{ler:pog:26}]\label{lem:weak_HPAL}
    The rules of internal and external weakening are height-preserving admissible in $\HPAL$.
    $$
    \vlinf{\weakr}{}{\hsq \hsep X \dsq \Gamma,A}{\hsq \hsep X \dsq \Gamma}
    \qquad
    \vlinf{\extweakr}{}{\hsq \hsep X \hsep Y}{\hsq \hsep X}
    $$
\end{lemma}

Finally, although not stated in \cite{ler:pog:26}, we further need to show that the so-called \emph{lift rule} is admissible in $\HPAL$. For defining this rule, let $C$ be a formula and $X$ be a dynamic hypersequent. The dynamic sequent $X\lift$ is obtained by replacing all indices $\ann$ occurring in $X$ by $C\cdot \ann$. In particular, if $X = \dsqq{\ann_1}\Gamma_1 \dsqq{}\cdots \dsqq{\ann_n}\Gamma_n$, the resulting dynamic sequent is $X\lift := \dsqq{C \cdot \ann_1}\Gamma_1 \dsqq{}\cdots \dsqq{C\cdot \ann_n}\Gamma_n$.
If $\hsq= X_1 \hsep \cdots \hsep X_k$, the dynamic hypersequent $\hsq\lift$ is obtained by lifting all dynamic sequents $X_i$ occurring in it, i.e., $\hsq\lift := X_1\lift \hsep \cdots \hsep X_k\lift$. Now, the \emph{lift rule} is:
\begin{equation}
  \label{eq:lift}
    \vlinf{\liftr}{}{\hsq\lift}{\hsq}
\end{equation}

\begin{lemma}\label{lem:lift_HPAL}
    The lift rule is height-preserving admissible in $\HPAL$.
\end{lemma}

\begin{proof}
    The proof proceeds by induction on the height of a derivation of the premise.
    If $\hsq$ is an axiom, then so is $\hsq\lift$.
    Otherwise, we consider the last applied rule $\ruler$ in the derivation: we apply the induction hypothesis to the premise(s) of $\ruler$ and then apply $\ruler$ again. This is straightforward. For the sake of clarity, the case where the last rule is an instance of $\pboxzr$ is provided below.
    $$
    \vlderivation{
    \vlin{\pboxzr}{}{\hsq \hsep X \dsq \PBOX AB,\Gamma}
    {\vlhtr{\hderi}{\hsq \hsep X \dsq A,\Gamma \dsqq{\ann \cdot A} B}}
    }
    \quad
    \byIH
    \quad
    \vlderivation{
    \vlin{\pboxzr}{}{\hsq\lift \hsep X\lift \dsqq{C \cdot \ann} \PBOX AB,\Gamma}
    {\vlhtr{\hderi'}{\hsq\lift \hsep X\lift \dsqq{C \cdot \ann} A,\Gamma \dsqq{C \cdot \ann \cdot A} B}}
    }
    $$
The other cases are similar. \qed
\end{proof}

\begin{remark}
    The admissibility of the lift rule allows to show that necessitation for announcements, as in the alternative axiomatisation provided in \cite{wang2013axiomatizations} is admissible in $\HPAL$. Since $\OPAL$ allows the construction of a derivation $\PBOX C\deri$ of $\PBOX CA$ from a derivation $\deri$ of $A$, it is not surprising that one needs to simulate the necessitation of announcement to show that $\OPAL$ is sound.
\end{remark}

\ThmOpalToHpal*

\begin{proof}
    Let $\odv{A}{\deri}{B}{\OPAL}$ be a derivation from $A$ to $B$ in $\OPAL$. In the following we show that $\proves[\HPAL+\cutr]{\cneg A,B}$ (we then get $\proves[\HPAL+\cutr]{\cneg A\OR B}$ by applying $\orr$, and $\proves[\PAL]{A \IMP B}$ follows from \Cref{thm:hpalcut-to-hpal} and \Cref{thm:HPAL}).
    
    We proceed by induction on the derivation $\deri$, and we have the following cases:

    \renewcommand{\labelenumi}{(\roman{enumi})}
\begin{enumerate}
\item     If $\deri$ is a formula, then $\deri =\odf{A}$ and $\proves[\HPAL]{\cneg A, A}$ by \Cref{lem:generalised_axioms}.
    \item Suppose $\deri=\odf{\odv{A}{\derib}{B}{}\ANDOR\odv{C}{\deric}{D}{}}$ where $\ANDOR \in \set{\AND,\OR}$. We only show the case where $\ANDOR = \AND$ (the case of $\OR$ is treated similarly), for which we need to show $\proves[\HPAL]{\cneg A \OR \cneg C, B \AND D}$.
    By induction hypothesis $\proves[\HPAL]{\cneg A, B}$ and $\proves[\HPAL]{\cneg C, D}$. By \Cref{lem:weak_HPAL} we can use weakening, and obtain the desired result by the following derivation.
    $$
    \vlderivation{
    \vlin{\orr}{}{\cneg A \OR \cneg C,B \AND D}{
    \vliin{\andr}{}{\cneg A, \cneg C, B \AND D}
    {\vlin{\weakr}{}{\cneg A, \cneg C, B}{\vlhtr{\hderi_1}{\cneg A, B}}}
    {\vlin{\weakr}{}{\cneg A,\cneg C,D}{\vlhtr{\hderi_2}{ \cneg C, D}}}}
    }
    $$
    
    \item Suppose $\deri=\odf{\BOXDIA\odv{A}{\derib}{B}{}}$ where $\BOXDIA \in \set{\BOX,\DIA,\PBOX C,\PDIA C}$.    
    We first show the case where $\BOXDIA = \BOX$ and then the case where $\BOXDIA = \PBOX C $ (those with $\DIA$ and $\PDIA C$ are treated similarly).
    
    If $\deri=\odf{\BOX\odv{A}{\derib}{B}{}}$, we need to show $\proves[\HPAL]{\DIA \cneg A, \BOX B}$. By induction hypothesis $\proves[\HPAL]{\cneg A,B}$.
    By \Cref{lem:weak_HPAL} we can use external weakening, and obtain the desired result by the following derivation.
    $$
    \vlderivation{
    \vlin{\boxr}{}{\DIA \cneg A, \BOX B}
    {\vlin{\diafr}{}{\DIA \cneg A \hsep B}{
    \vlin{\extweakr}{}{\DIA \cneg A \hsep \cneg A,B}{\vlhtr{\hderi}{\cneg A,B}}
    }}}
    $$

    If $\deri=\odf{\PBOX C\odv{A}{\derib}{B}{}}$, we need to show $\proves[\HPAL]{\PDIA C \cneg A, \PBOX C B}$. By induction hypothesis $\proves[\HPAL]{\cneg A,B}$.
    Here we use \Cref{lem:lift_HPAL} together with \Cref{lem:weak_HPAL} and reach the desired result by the following derivation, where $\hderi_1$ exists by \Cref{lem:generalised_axioms} and $\hderi_2$ exists by induction hypothesis.
    $$
    \vlderivation{
    \vlin{\pboxzr}{}{\PDIA C \cneg A,\PBOX CB}{
    \vliin{\pdiar}{}{\cneg C,\PDIA C\cneg A \dsqq{C} B}
    {\vlhtr{\hderi_1}{C, \cneg C \dsqq{C}B}}
    {\vlin{\weakr}{}{\cneg C \dsqq{C}\cneg A,B}{
    \vlin{\liftr}{}{\dsqq{C} \; \cneg A,B}{\vlhtr{\hderi_2}{\cneg A,B}}
    }}}}
    $$

    \item Finally, suppose $\deri=\odn{\odv{A}{\derib}{B}{}}{\rrule}{\odv{C}{\deric}{D}{}}{}$ where $\vlinf{\rrule}{}{C}{B}$ is an instance of an inference rule. We need to show $\proves[\HPAL]{\cneg A, D}$.
    By induction hypothesis we have $\proves[\HPAL]{\cneg A, B}$ and $\proves[\HPAL]{\cneg C, D}$. Moreover, by \Cref{lem:rule-sound} we have $\proves[\HPAL]{\cneg B, C}$. By \Cref{thm:cut_HPAL} we can use $\cutr$ twice to obtain the desired conclusion as follows:
    $$
    \vlderivation{
    \vliin{\cutr}{}{\cneg A,D}
    {\vlhtr{\hderi_1}{\cneg A, B}}{\vliin{\cutr}{}{\cneg B,D}{\vlhtr{\hderi_2}{\cneg B, C}}{\vlhtr{\hderi_3}{\cneg C, D}}}
    }
    $$
\end{enumerate}
\qed
\end{proof}


\section{Missing proof for \Cref{sec:completeness}}\label{app:completeness}

\LemHpalOpal*

\begin{proof}
  We construct $\deri$ for each rule of $\HPAL$.
  \begin{itemize}
  \item For $\topr$, we use for $\deri$ the base case of \Cref{lem:and-n}.
  \item For $\axr$, we have $\deri$ as follows:
    $$
    \vlinf{\axr}{}{\hsq\hsep X\dsq \Gamma,\cneg \atom,\atom}{}
    \qualto
    \odv{\top}{\deri'}{\intof\hsq\OR\BOX(\intof X\OR\PBOX\ann(\intof\Gamma\OR
      \odn{\top}{\aiD}{\cneg \atom\OR \atom}{}))}{}
    $$
    where $\deri'$ is again the base case of the proof of \Cref{lem:and-n}.
  \item For $\orr$, $\deri$ is the identity derivation:
    $$
    \vlinf{\orr}{}{\hsq\hsep X\dsq\Gamma, A\OR B}{\hsq\hsep X\dsq\Gamma, A, B}
    \qualto
    \odf{\intof\hsq\OR\BOX(\intof X\OR\PBOX{\ann}(\intof\Gamma\OR(A\OR B)))}
    $$
  \item For $\andr$, $\deri$ directly obtained from \Cref{lem:and-2}:
    $$
    \vliinf{\andr}{}{\hsq\hsep X\dsq\Gamma, A\AND B}{
      \hsq\hsep X\dsq\Gamma, A}{
      \hsq\hsep X\dsq\Gamma, B}
    \qualto
    \odv{\intof\hsq\OR\BOX(\intof X\OR\PBOX{\ann}(\intof\Gamma\OR A))\AND\intof\hsq\OR\BOX(\intof X\OR\PBOX{\ann}(\intof\Gamma\OR B))}{\deri}{
      \intof\hsq\OR\BOX(\intof X\OR\PBOX{\ann}(\intof\Gamma\OR (A\AND B)))}{}
    $$
  \item For $\boxr$ we have
    $$
    \vlinf{\boxr}{}{\hsq\hsep X\dsq\Gamma, \BOX A}{
    \hsq\hsep X\dsq\Gamma \ \hsep \dsq \  A}
    \qualto
    \intof{\hsq} \OR
    \odn{\BOX( \intof{X} \OR \PBOX \alpha \intof{\Gamma}) \OR
      \BOX\PBOX \alpha A
    }
        {\medvD}{\BOX\left(
          \intof{X} \OR
          \odn{\PBOX \alpha \intof{\Gamma} \OR
            \odn{\BOX \PBOX \alpha A}{\boxhPD}{\PBOX \alpha \BOX A}{}}
              {\medhPD}{\PBOX \alpha(\intof{\Gamma} \OR \BOX A)}{}
              \right)}{}
        $$
  \item For $\diatr$, we have
    $$
    \vlinf{\diatr}{}{\hsq\hsep X\dsq\Gamma, \DIA A}{
      \hsq\hsep X\dsq\Gamma, \DIA A, A}
    \qualto
    \intof{\hsq} \OR \BOX\left( \intof{X} \OR \PBOX \alpha \left(
    \intof{\Gamma} \OR
    \odn{ \DIA A \OR \odn{A}{\tD}{\DIA A}{}}{\conD}{\DIA A}{} \right) \right)
    $$
  \item For ~$\vlinf{\diafr}{}{\hsq\hsep X\dsq\Gamma, \DIA A\hsep Y\dsq\Delta}{
    \hsq\hsep X\dsq\Gamma, \DIA A\hsep Y\dsq\Delta,A}$~ we have
    $$
    \intof\hsq\OR
    \odn{\BOX(\intof X\OR\PBOX\ann(\intof\Gamma\OR\DIA A))
      \OR
      \BOXof{\intof Y\OR\odn{\PBOX\ann(\intof \Delta\OR A)}{
        \khPD}{\PBOX\ann\intof\Delta\OR\PDIA\ann A}{}}
    }{\kfD}{
      \BOXof{
        \intof X\OR
        \odn{\PBOX\ann(\intof \Gamma\OR\DIA A)\OR
        \odn{\DIA\PDIA\ann A}{\diahPD}{\PBOX\ann\DIA A}{}}{
          \medhPD}{
        \PBOXof\ann{\intof\Gamma\OR\odn{\DIA A\OR\DIA A}{\conD}{\DIA A}{}}}{}}
      \OR\BOX(\intof Y\OR\PBOX\ann\intof\Delta)
    }{}
    $$
  \item For ~$\vliiiiinf{\diapr}{}{\hsq\hsep X \dsqq{\anna\acomp\annb} \Gamma, \DIA A
        \hsep Y\dsq\Delta}{
        \hsqq{G'}\hsep Y\dsq\Delta,B_1}{
        ~\hsqq{G'}\hsep Y\dsq\Delta\dsqq{\ann\acomp B_1}B_2}{
        ~\hsqq{G'}\hsep Y\dsq\Delta\dsqq{\ann\acomp B_1\acomp B_2}B_3}{
        \cdots}{
        \hsqq{G'}\hsep Y\dsq\Delta\dsqq{\ann\acomp\annb}A}
    $ we have
    $$
    \scalebox{.85}{$
    \odn{
      \odv{G\OR\BOXof{\intof Y\OR\PBOXof\ann{\intof\Delta\OR B_1}})
        \AND
        \left(G\OR\BOXof{\intof Y\OR
          \odn{\PBOX\ann\intof\Delta\OR\PBOX{\ann.B_1}B_2}{
            \medhPD}{
            \PBOXof\ann{\intof\Delta\OR \PBOX{B_1}B_2}}{}}\right)
        \AND\cdots\AND
        \left(G\OR\BOXof{\intof Y\OR
          \odn{\PBOX\ann\intof\Delta\OR\PBOX{\ann \cdot \annb}A}{
            \medhPD}{
            \PBOXof\ann{\intof\Delta\OR \PBOX{\annb}A}}{}}\right)
      }{
        \deri_1}{
        G\OR\BOXof{\intof Y\OR\PBOXof\ann{\intof\Delta\OR
            \odv{B_1\AND\PBOX{B_1}B_2\AND\PBOX{B_1.B_2}B_3
              \AND\cdots\AND\PBOX{\annb}A}{\deri_0}{
              \PDIA\annb A}{}}}}{}}{
      \equiv}{
        \intof\hsq\OR
        \odn{\BOX(\intof X\OR\PBOX{\ann \cdot \annb}(\intof\Gamma\OR\DIA A))
          \OR
          \BOXof{\intof Y\OR\odn{\PBOX{\ann}(\intof \Delta\OR\PDIA{\annb} A)}{
              \khPD}{\PBOX{\ann}\intof\Delta\OR\PDIA{\ann \cdot \annb} A}{}}
    }{\kfD}{
      \BOXof{
        \intof X\OR
        \odn{\PBOX{\ann \cdot \annb}(\intof \Gamma\OR\DIA A)\OR
        \odn{\DIA\PDIA{\ann \cdot \annb} A}{\diahPD}{\PBOX{\ann \cdot \annb}\DIA A}{}}{
          \medhPD}{
        \PBOXof{\ann \cdot \annb}{\intof\Gamma\OR\odn{\DIA A\OR\DIA A}{\conD}{\DIA A}{}}}{}}
      \OR\BOX(\intof Y\OR\PBOX{\ann}\intof\Delta)
        }{}}{}
    $}
    $$
    where
    $G=\intof{(\hsq')}=\intof{(\hsq\hsep X \dsqq{\anna\acomp\annb} \Gamma, \DIA A)}
    =\intof\hsq\OR\BOX(\intof X\OR\PBOX{\ann\acomp\annb}(\intof\Gamma\OR\DIA A))$,
    and $\deri_1$ is obtained from \Cref{lem:and-n}, and $\deri_0$ is the derivation
    $$
    \od{\odI{\odI{\odh{B_1\AND\PBOX{B_1}B_2\AND\PBOX{B_1 \cdot B_2}B_3
            \AND\cdots\AND\PBOX{B_1 \cdot B_2 \cdots B_n}A}}{
          (n-1)\times\medPD}{
        B_1\AND\PBOX{B_1}(B_2\AND\PBOX{B_2}(B_3
        \AND\cdots\PBOX{B_{n-1}}(B_n\AND\PBOX{B_n}A)\cdots))}{}}{
        (n-1)\times\redD}{
        \PDIA{B_1 \cdot B_2 \cdots B_n}A}{}}
    $$    
  \item for $\pboxzr$, we have
    $$
    \vlinf{\pboxzr}{}{\hsq\hsep X\dsq\Gamma,\PBOX AB}{
      \hsq\hsep X\dsq\Gamma, \cneg A\dsqq{\ann\acomp A}B}
    \qualto
    \intof\hsq\OR\BOXof{
      \intof X\OR
      \odn{\PBOX\ann(\intof\Gamma\OR\cneg A)\OR\PBOX{\ann\acomp A}B}{
        \medhPD}{
        \PBOXof{\ann}{\intof\Gamma\OR
          \odn{\odn{\cneg A}{\negD}{\PBOX AB}{}\OR\PBOX AB}{\conD}{\PBOX AB}{}}
      }{}}
    $$
  \item for $\vlinf{\pboxr}{}{\hsq\hsep X\dsq\Gamma,\PBOX AB\dsqq{\ann\acomp A}\Delta}{
    \hsq\hsep X\dsq\Gamma, \cneg A\dsqq{\ann\acomp A}\Delta,B}
    $\; we have
    $$
    \intof\hsq\OR\BOXof{
      \intof X\OR
      \odn{\PBOX\ann(\intof\Gamma\OR\cneg A)\OR
        \odn{\PBOX{\ann\acomp A}(\intof\Delta\OR B)}{
          \khPD}{
          \PBOX{\ann\acomp A}\intof\Delta\OR\PDIA{\ann\acomp A}B}{}
      }{\equiv}{
      \odn{\PBOX\ann(\intof\Gamma\OR\cneg A)\OR
        \odn{\PDIA{\ann\acomp A}B}{\funhPD}{\PBOX{\ann\acomp A}B}{}}{
        \medhPD}{
        \PBOXof{\ann}{\intof\Gamma\OR
          \odn{\odn{\cneg A}{\negD}{\PBOX AB}{}\OR\PBOX AB}{\conD}{\PBOX AB}{}}
      }{}
      \OR \PBOX{\ann\acomp A}\intof\Delta
      }{}}
    $$
   
  \item for $\vliinf{\pdiazr}{}{\hsq\hsep X\dsq\Gamma,\PDIA AB}{
      \hsq\hsep X\dsq\Gamma, A}{
      \hsq\hsep X\dsq\Gamma\dsqq{\ann\acomp A} B}$ we have
    $$
    \odv{
      (\intof\hsq\OR\BOX(\intof X\OR
      \PBOX\ann(\intof\Gamma\OR A)))
      \AND
      \left(\intof\hsq\OR\BOXof{\intof X\OR
        \odn{\PBOX\ann\intof\Gamma\OR\PBOX{\ann.A}B}{
          \medhPD}{\PBOX\ann(\intof\Gamma\OR\PBOX{A}B)}{}
      }\right)}{
      \deri'}{
      \intof\hsq\OR\BOXof{\intof X\OR
      \PBOXof\ann{\intof\Gamma\OR
      \odn{A\AND\PBOX AB}{\redD}{\PDIA AB}{}}}}{}
    $$
    where $\deri'$ exists by \Cref{lem:and-2}.
  \item For $\vliinf{\pdiar}{}{\hsq\hsep X\dsq\Gamma,\PDIA AB\dsqq{\ann\acomp A}\Delta}{
      \hsq\hsep X\dsq\Gamma, A\dsqq{\ann\acomp A}\Delta}{
      \hsq\hsep X\dsq\Gamma\dsqq{\ann\acomp A}\Delta, B}
    $\; we have
    $$
    \odv{
      (\intof\hsq\OR\BOX(\intof X\OR
      \PBOX\ann(\intof\Gamma\OR A)\OR\PBOX{\ann\acomp A}\intof\Delta))
      \AND
      \left(\intof\hsq\OR\BOXof{\intof X\OR
        \odn{\PBOX\ann\intof\Gamma\OR
          \odn{\PBOX{\ann\acomp A}(\intof\Delta\OR B)}{\kPD}{\PBOX{\ann\acomp A}\intof\Delta\OR\PDIA{\ann\acomp A}B}{}
        }{\equiv}{
          \odn{\PBOX\ann\intof\Gamma\OR
            \odn{\PDIA{\ann\acomp A}B}{\funhPD}{\PBOX{\ann\acomp A}B}{}
          }{
            \medhPD}{\PBOX\ann(\intof\Gamma\OR\PBOX{A}B)}{}
          \OR\PBOX{\ann\acomp A}\intof\Delta}{}
        }\right)}{
      \deri'}{
      \intof\hsq\OR\BOXof{\intof X\OR
      \PBOXof\ann{\intof\Gamma\OR
      \odn{A\AND\PBOX AB}{\redD}{\PDIA AB}{}}\OR\PBOX{\ann\acomp A}\intof\Delta}}{}
    $$
    where $\deri'$ exists by \Cref{lem:and-2}.
  \item For ~$\vlinf{\atr}{}{\hsq\hsep X\dsq\Gamma\dsqq{\ann\acomp A}\Delta,\atomb}{
      \hsq\hsep X\dsq\Gamma, \atomb\dsqq{\ann\acomp A}\Delta,\atomb}
    $\; we construct
    $$
    \intof\hsq\OR\BOXof{\intof X\OR
      \odn{\odn{\PBOXof\ann{\intof\Gamma\OR\odn{\atomb}{\necPD}{\PBOX A\atomb}{}}}{
          \khPD}{
          \PBOX\ann\intof\Gamma\OR\PDIA\ann\PBOX A\atomb}{}
        \OR\PBOX{\ann\acomp A}(\intof\Delta\OR \atomb)}{
        \equiv}{
        \PBOX\ann\intof\Gamma\OR
        \odn{\PBOX{\ann\acomp A}(\intof\Delta\OR \atomb)\OR
          \odn{\PDIA\ann\PBOX{A} \atomb}{\funhPD}{\PBOX{\ann\acomp A} \atomb}{}}{
          \medhPD}{
          \PBOX{\ann\acomp A}(\intof\Delta\OR \odn{\atomb\OR \atomb}{\conD}{\atomb}{})}{}
      }{}
    }
    $$
  \item For $\newr$, we have
    $$
    \vlinf{\newr}{}{\hsq\hsep X\dsqq{\ann\acomp A}\Delta}{
      \hsq\hsep X\dsq{\!}\dsqq{\ann\acomp A}\Delta}
    \qualto
    \intof\hsq\OR \BOXof{\intof X \OR
      \odn{\PBOX \ann \bot\OR \PBOX {\ann\acomp A}\intof\Delta}{
        \medhPD}{
        \PBOXof \ann {
          \odn{\bot \OR \PBOX{A}\intof\Delta }{
          \equiv}{
          \PBOX{A}\intof\Delta 
        }{}}}{}}
    $$
  \item For $
    \vlinf{\recallr}{}{\hsq\hsep X\dsq\Gamma\dsqq{\ann\acomp A}\Delta}{
      \hsq\hsep X\dsq\Gamma, \cneg A\dsqq{\ann\acomp A}\Delta}
    $ we have
    $$
    \intof\hsq\OR \BOXof{\intof X \OR
      \odn{\odn{\PBOX\ann(\intof\Gamma\OR\cneg A)}{
          \khPD}{
          \PBOX\ann{\intof\Gamma}\OR\PDIA \ann \cneg A}{}
        \OR \PBOX {\ann \acomp A}\intof\Delta}{
        \equiv}{
        \PBOX\ann{\intof\Gamma}\OR
        \odn{\odn{\PDIA \ann \cneg A}{\funhPD}{\PBOX \ann \cneg A}{}\OR \PBOX {\ann\acomp A}\intof\Delta}{
          \medhPD}{
          \PBOXof \ann {
            \odn{\odn{\cneg A}{\negD}{\PBOX{A}\intof\Delta}{}
              \OR \PBOX{A}\intof\Delta }{
              \conD}{
              \PBOX{A}\intof\Delta 
            }{}}}{}}{}}    
    $$
  \qed
  \end{itemize}
\end{proof}

\end{document}